\documentclass{IEEEtran}
\usepackage{cite}
\usepackage{amsmath,amssymb,amsfonts}
\usepackage{mathrsfs}
\usepackage{algorithmic}
\usepackage{graphicx}
\usepackage{textcomp}
\usepackage{xcolor}
\usepackage{changepage} 
\usepackage{setspace}
\usepackage{cases}  
\usepackage{arydshln}  
\usepackage{bm}  

\newtheorem{lemma}{Lemma}
\newtheorem{remark}{Remark}
\newtheorem{definition}{Definition}
\newtheorem{theorem}{Theorem}

\newtheorem{proposition}{Proposition}
\newtheorem{assumption}{Assumption}
\newcommand{\pfbox}{\hfill\mbox{$\Box$}}
\newcounter{MYtempeqncnt} 
\usepackage{dblfloatfix} 

\def\BibTeX{{\rm B\kern-.05em{\sc i\kern-.025em b}\kern-.08em
		T\kern-.1667em\lower.7ex\hbox{E}\kern-.125emX}}
\begin{document}
	\title{\bf{Multi-Objective Complementary Control}}
    
	\author{Jiapeng Xu, Xiang Chen, Ying Tan, and Kemin Zhou
		\thanks{Jiapeng Xu and Xiang Chen are with the Department of
			Electrical and Computer Engineering,
			University of Windsor, 401 Sunset Avenue, Windsor, ON N9B 3P4, Canada (e-mails: imxjp@sina.com; xchen@uwindsor.ca).}
        \thanks{Ying Tan is with the Department of Mechanical Engineering, The
			University of Melbourne, Melbourne, VIC 3010, Australia (e-mail: yingt@unimelb.edu.au).}
		\thanks{Kemin Zhou is with the Center for Advanced Control and Smart Operations, Nanjing University, Suzhou, Jiangsu 215163,
			China (e-mail: kmzhou@nju.edu.cn).}}
   
	\maketitle
	
		\begin{abstract}
		This paper proposes a novel multi-objective control framework for linear time-invariant systems in which performance and robustness can be achieved in a complementary way instead of a trade-off. 
        In particular, a state-space solution is first established for a new stabilizing control structure consisting of two independently designed controllers coordinated with a Youla-type operator ${\bm Q}$. It is then shown by performance analysis that these two independently designed controllers operate in a naturally complementary way for a tracking control system, due to the coordination function of ${\bm Q}$ driven by the residual signal of a Luenberger observer.  Moreover, it is pointed out that ${\bm Q}$ could be further optimized with an additional gain factor to achieve improved performance, through a data-driven methodology 
  for a measured cost function. 
	\end{abstract}
	
	
	\section{Introduction}
	There are usually multiple objectives or specifications required to be achieved by various control systems in practice \cite{khargonekar1991multiple,scherer1997multiobjective,elia1997controller,chen2001multiobjective,lin2013achieving,menini2018algebraic,balandin2019multi,bhowmick2020solution,han2023multiobjective}, such as optimality of tracking performance, robustness to some unknown/uncertain disturbances or parameter variations, passivity, etc.
	Various formulated multi-objective (MO) control problems have received considerable attention in the control community, for example, the MO $\mathcal{H}_{2}/\mathcal{H}_{\infty}$ control \cite{bernstein1989lqg,khargonekar1991mixed,limebeer1994nash,zhou1994mixed,doyle1994mixed,hindi1998multiobjective,chen2001multiobjective}, where optimization of performance in the $\mathcal{H}_2$ norm and robustness in the $\mathcal{H}_{\infty}$ norm are considered simultaneously and are heavily studied in 1990's. Other kinds of MO optimal control problems have also been extensively explored, for example, in \cite{khargonekar1991multiple,elia1997controller,hindi1998multiobjective,balandin2019multi,menini2018algebraic}, where the concept of Pareto optimality is used with respect to multiple optimization criteria.
	
	It can be observed that the common feature of these traditional MO control problems is the fact that a single-controller structure is applied which poses challenges for the design methodologies, especially, when the involved objectives are inherently conflicting. A typical example is the well-known conflicting pair of robustness and optimal performance in a traditional robust control design. When conflicting objectives could not be achieved at their best by the single controller, the trade-off is normally the only choice, which results in a compromised single controller, as can be seen in many mixed $\mathcal{H}_{2}/\mathcal{H}_{\infty}$ control results \cite{bernstein1989lqg,khargonekar1991mixed,limebeer1994nash,doyle1994mixed,hindi1998multiobjective,chen2001multiobjective}. There are some research reported for controller designs in the two-degree-of-freedom (2DoF) form \cite{youla1985feedback,moore1986improving,wen2022bi}.  The reference \cite{youla1985feedback} is concerned with 2DoF optimal design for a quadratic cost. In particular, the class of all stabilizing 2DoF controllers which give finite cost is characterized. In \cite{moore1986improving}, the class of all stabilizing 2DoF controllers which achieve a specified closed-loop transfer matrix is characterized in terms of a free stable parameter. Inspired by the generalized internal model control (GIMC) developed in \cite{zhou2001new}, a bi-objective high-performance control design for antenna servo systems is studied in \cite{wen2022bi}. Performance limitation problems are also studied for 2DoF controllers in \cite{chen2003best,chen2000limitations}, with tracking and/or regulation as a sole objective. However, there is, essentially, a lack of a fundamental and general framework that can assemble two controllers together in a complementary way for MO control design purposes. That said, it is desired to develop a new framework for MO control problems that can overcome the curse of trade-offs and achieve non-compromised MO performances, which is the main goal of this paper.
	
	This work is closely related to the GIMC structure \cite{zhou2001new}, a multi-block implementation of the famous Youla-Ku\v cera parameterization of all stabilizing  1DoF and 2DoF controllers \cite{vidyasagar2022control}. In this implementation, the Youla-type parameter $\bm{Q}(s)$ becomes an explicit design factor driven by the residual signal, instead of functioning as an optional parameter to deliver a specific stabilizing controller. Although this GIMC structure provides hope for the desired two-controller complementary structure to address MO control problems, no further details are given for the systematic design of $\bm{Q}(s)$ in \cite{zhou2001new}. Motivated by the GIMC, in this paper a new two-controller design framework called `Multi-Objective Complementary Control' (MOCC) is proposed and explored in detail with rigorous performance analysis, aiming to achieve MO performances while alleviating trade-offs. 
    In particular, state-space formulas are provided for the MOCC framework.
    A tracking control setting is utilized as a platform to demonstrate the said advantages of MOCC, that is, overcoming the trade-off in the conventional MO control to achieve nominal tracking performance and robustness in a complementary way. 
	Furthermore, a data-driven optimization approach is also sketched for the design of Youla-type operator $\bm{Q}$ to turn a robust controller into an optimal one, when performing the tracking task repetitively over a finite time interval in the presence of an unknown but repetition-invariant disturbance.

	This paper is organized as follows: in Section \ref{sec:preliminaries}, the motivation of this paper is explained in detail using a tracking control case as well as some preliminaries of GIMC and Youla-Ku\v cera parameterization; Section \ref{sec:tcs} presents a new two-controller structure which enables the MOCC design in the later section; in Section \ref{sec:MOCC}, a tracking control problem is addressed in the new framework with a rigorous performance analysis, illustrating the design features and the advantages of MOCC; the data-driven performance optimization is sketched in Section \ref{sec:optimization}; simulation results can be found in Section \ref{sec:example} and conclusions in Section \ref{sec:conclusion}.

	\emph{Notations:} Throughout this paper, the symbol of transfer matrices or systems will be in bold to distinguish them from constant matrices. A system and its transfer function are denoted by the same symbol and whenever convenient, the dependence on the frequency variable $s$ or $j\omega$ for a transfer function may be omitted.
	The set $\mathbb{R}^{n}$ consists of all $n$-dimensional real vectors. The unsubscripted norm $\|\cdot\|$ denotes the standard Euclidean norm on vectors.
	For a matrix or vector $X$, $X'$ denotes its transpose. For a rational transfer matrix $\bm{T}(s)$, $\bm{T}^*$ denotes $\bm{T}'(-s)$ or complex conjugation of $\bm{T}(j\omega)$.
    $\mathcal{RL}_{\infty}$ ($\mathcal{RH}_{\infty}$) denotes the space of all rational (and stable) functions with the norm $\displaystyle{\|\bm{T}(s)\|_{\infty}=\sup_{\omega}\bar\sigma\{\bm{T}(j\omega)\}}$ for any $\bm{T}(s)\in\mathcal{RL}_{\infty}(\mathcal{RH}_{\infty})$, where $\bar\sigma\{\cdot\}$ represents the largest singular value.
	A rational proper transfer matrix in terms of state-space data is simply denoted by $\left[\begin{array}{c|c}
		A & B \\
		\hline
		C & D
	\end{array}\right]:=C(sI-A)^{-1}B+D$. 
		
	\section{Motivation}\label{sec:preliminaries}
\begin{figure}[!t]
			\centering
			\includegraphics[scale=0.5]{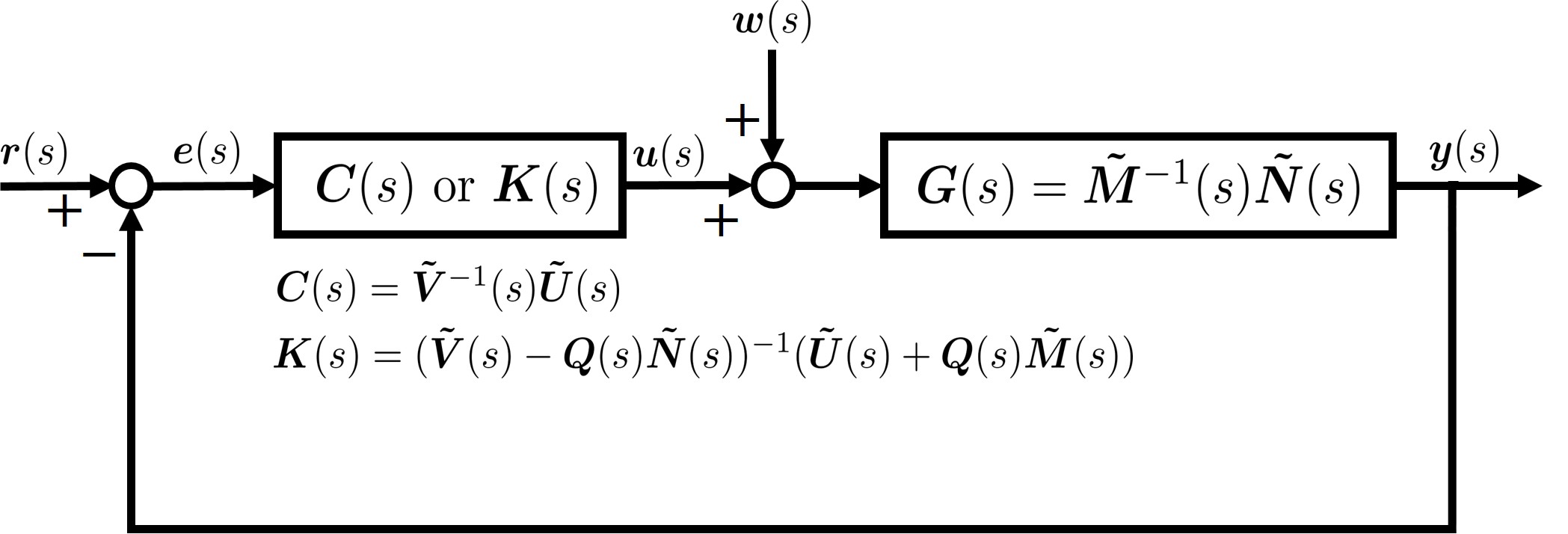}
			\caption{Tracking control system.}
			\label{fig:feedback_disturbance}
		\end{figure}

        Transfer functions shall be utilized to motivate our idea, though this paper mainly works on state space. Consider a tracking control system shown in Fig. \ref{fig:feedback_disturbance}. Let $\bm{G}(s)=\bm{\tilde M}^{-1}(s)\bm{\tilde N}(s)$ be a linear time-invariant (LTI) plant with a left coprime factorization $\bm{\tilde M}(s)\in{\cal RH}_{\infty}$ and $\bm{\tilde N}(s)\in{\cal RH}_{\infty}$, $\bm{C}(s)=\bm{\tilde V}^{-1}(s)\bm{\tilde U}(s)$ be a stabilizing controller with a left coprime factorization $\bm{\tilde V}(s)\in{\cal RH}_{\infty}$ and $\bm{\tilde U}(s)\in{\cal RH}_{\infty}$, $\bm{r}(s)$ be a reference signal to be tracked, and $\bm w(s)$ be an uncertain or unknown disturbance signal. 
         It is well-known that all stabilizing controllers can be characterized in the format of Youla-Ku\v cera parameterization 
        \begin{align*}
            \bm{K}=(\bm{\tilde V}-\bm{Q}\bm{\tilde N})^{-1}(\bm{\tilde U}+\bm{Q}\bm{\tilde M})
        \end{align*}
         for the same plant $\bm{G}(s)$, with $\bm{Q}(s)\in{\cal RH}_{\infty}$ and ${\rm det} (\bm{\tilde V}(\infty)-\bm{Q}(\infty)\bm{\tilde N}(\infty))\neq 0$ \cite{zhou2001new}, \cite[Chapter 5]{vidyasagar2022control}, \cite{anderson1998youla}.	 The tracking error $\bm{e}(s)=\bm{r}(s)-\bm{y}(s)$ with either $\bm{C}(s)$ or $\bm{K}(s)$ is given by
		\begin{align*}
			\bm{e}&=(I+\bm{G}\bm{C})^{-1}\bm{r}-(I+\bm{G}\bm{C})^{-1}\bm{G}\bm{w}, \\
			\text{or} \; \bm{e}&=(I+\bm{G}\bm{K})^{-1}\bm{r}-(I+\bm{G}\bm{K})^{-1}\bm{G}\bm{w}.
		\end{align*}
It can be seen that the single controller $\bm{C}(s)$ (or $\bm{K}(s)$) has to be designed to deal with both the tracking performance for the reference signal $\bm{r}(s)$ and the robustness against the disturbance $\bm{w}(s)$ simultaneously. In practice, trade-off usually should be carried out to compromise the design difficulty of the single $\bm{C}(s)$ (or $\bm{K}(s)$) between the tracking performance and the disturbance attenuation.  
On the other hand, considering the challenge posed for the selection of $\bm{Q}(s)$ under a given performance expectation of $\bm{K}(s)$, one can conclude that finding a $\bm{K}(s)$ is not necessarily easier than designing $\bm{C}(s)$ when both performance and robustness need to be addressed. 
		
\begin{figure}[!t]
			\centering
			\includegraphics[scale=0.45]{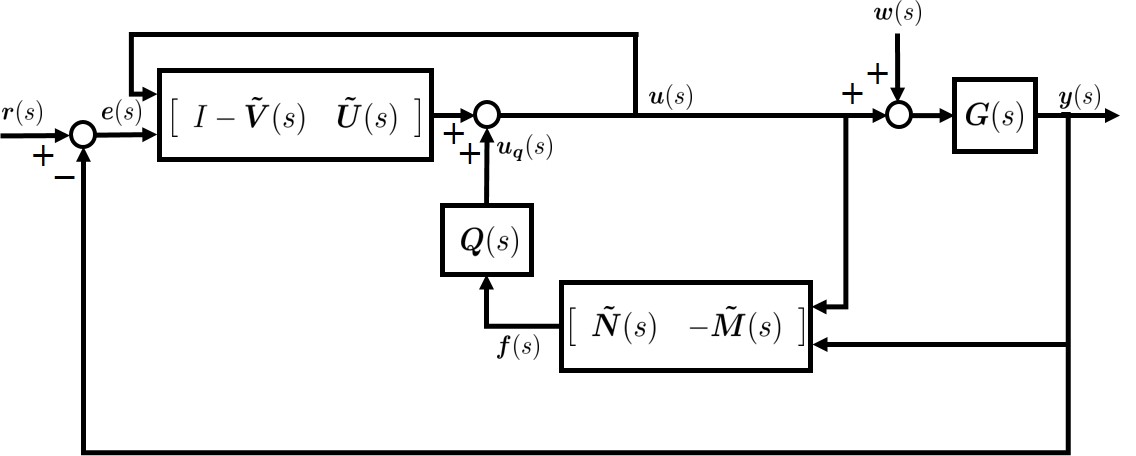}
			\caption{Generalized internal model control (GIMC) structure.}
			\label{fig:GIMC_disturbance}
		\end{figure}

    An interesting controller architecture called GIMC was proposed in \cite{zhou2001new} which provides a promising approach to overcome the trade-off in the traditional design. The GIMC is shown in Fig. \ref{fig:GIMC_disturbance}, featuring the parameter $ \bm{Q}(s)\in{\cal RH}_{\infty}$ as a separate design factor in the structure, instead of being buried in $\bm{K}(s)$.
		%
		%
		%
		By direct algebra, 
        the tracking error $\bm{e}(s)$ in Fig. \ref{fig:GIMC_disturbance} can be derived as
		\begin{align*}
			\bm{e}=&(I+\bm{G}\bm{C})^{-1}\bm{r}-(I+\bm{G}\bm{C})^{-1}\bm{G}(I-\bm{\tilde V}^{-1}\bm{Q}\bm{\tilde N})\bm{w}.
		\end{align*}
		It is clear from the above expression that different from the control structure in Fig. \ref{fig:feedback_disturbance}, there are potentially two separate design degrees of freedom for tracking control in GIMC:  
		$\bm{C}(s)$ for tracking performance and $\bm{Q}(s)$ for disturbance attenuation. 
         In addition, if there is no uncertainty, i.e., $\bm w=0$, then $\bm{e}=(I+\bm{G}\bm{C})^{-1}\bm{r}$ and thus the nominal tracking error signal is recovered. These attractive features motivate us to revisit MO control problems through the design of $\bm{C}(s)$ and $\bm{Q}(s)$, instead of the traditional design of a single controller $\bm{C}(s)$ or $\bm{K}(s)$. Noting that, for a given $\bm{C}(s)$, if $\bm{Q}(s)$ can be determined then $\bm{K}(s)$ is derived and vice verse. Hence, design of $\bm{C}(s)$ and $\bm{Q}(s)$ can be reduced to the design of $\bm{C}(s)$ and $\bm{K}(s)$ even if $\bm{K}(s)$ is not explicitly seen in Fig \ref{fig:GIMC_disturbance}. It is noted that how to coordinate $\bm{C}(s)$ and $\bm{K}(s)$ using $\bm{Q}(s)$ such that specified performances are achieved and rigorous performance analysis for GIMC  have not systematically and profoundly been investigated so far.
        In \cite{chen2019revisit}, a solution to $\bm{Q}(s)$ is derived in state space for a robust LQG control problem through the $\mathcal{H}_{\infty}$ design with a modified control structure from GIMC, although it is still not clear how to systematically use $\bm{Q}(s)$ to coordinate $\bm{C}(s)$ and $\bm{K}(s)$ in the Youla-Ku\v cera parameterization. 

		In the following sections of this paper, a two-controller structure $\bm {K_{CQ}}$ is proposed  
        in state space, motivated from the analysis above. In particular, how $\bm Q$ can be constructed from two independently designed controllers $\bm C$ and $\bm K$ to coordinate $\bm C$ and $\bm K$ is explored systematically. To showcase the superior performance of this new control design framework, a tracking control problem is presented to demonstrate how $\bm C$ and $\bm K$ can be coordinated through $\bm Q$  to address MO tracking performance and robustness in a complementary way, together with a rigorous performance analysis.  

\begin{figure}[!ht]
			\centering
            \includegraphics[scale=0.45]{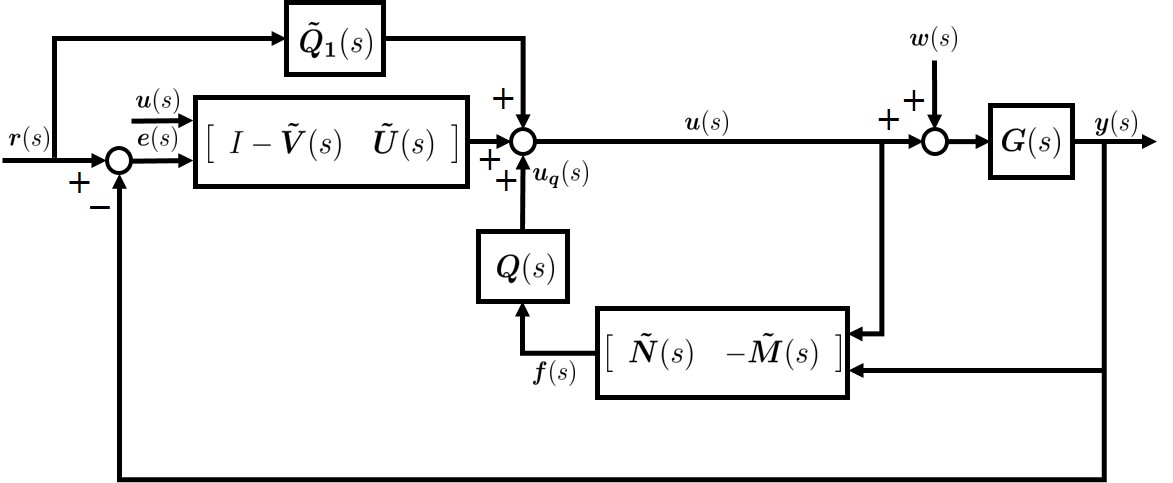}
			\caption{A GIMC implementation of 2DOF controller.}\label{fig:GIMC_2dof}
		\end{figure}

        \begin{remark}\label{rem:2dof}
            Note that the GIMC structure in Fig. \ref{fig:GIMC_disturbance} can be extended to an equivalent 2DoF controller structure with an additional 
            feed-forward parameter $\bm {\tilde Q_1}(s)\in\mathcal{RH_\infty}$, shown in Fig \ref{fig:GIMC_2dof}. Indeed, the controller in Fig. \ref{fig:GIMC_2dof} can be deemed as $\bm{u} = \bm{K_1r}-\bm{K_2y}$ with
            \begin{align*}
                \left[\begin{array}{cc}
					\bm {K_1} & \bm {K_2}
				\end{array}\right]=(\bm{\tilde V}-\bm{Q}\bm{\tilde N})^{-1} \left[\begin{array}{cc}
					\bm {\tilde U}+\bm {\tilde Q_1} & (\bm{\tilde U}+\bm{Q}\bm{\tilde M})
				\end{array}\right],
            \end{align*}
            which in fact parameterizes all stabilizing 2DoF controllers in terms of free parameters $\bm {\tilde Q_1}(s)\in\mathcal{RH_\infty}$ and $\bm{Q}(s)\in\mathcal{RH_\infty}$ \cite[Theorem 5.6.3]{vidyasagar2022control}. In this case, 
            the controller $\bm C(s)$ can be obtained as $\bm {\tilde V}^{-1}\left[\begin{array}{cc}
					\bm {\tilde U}+\bm {\tilde Q_1} & \bm {\tilde U}
				\end{array}\right]$. Since $\bm {\tilde Q_1}(s)\in\mathcal{RH_\infty}$ occurs in the feed-forward channel, without loss of generality, the GIMC in Fig. \ref{fig:GIMC_disturbance} is adopted as the beginning structure in this paper. \pfbox
        \end{remark}

		\section{Two-Controller Structure}\label{sec:tcs}
		
		This section presents a new stabilizing control structure motivated by GIMC. Consider the following finite-dimensional linear time-invariant (FDLTI) system
		\begin{align}\label{LTIsys:stabilizing}
			\bm{G}:\left\{ \mspace{-6mu}\begin{array}{l}
				\dot x=Ax+B_2u\\
				y=C_2x,
			\end{array} \right.
		\end{align}
		where $x\in\mathbb{R}^n$, $u\in\mathbb{R}^{m_2}$ and $y\in\mathbb{R}^{p_2}$ are the state, control input, and sensor output, respectively. All matrices are real constant and have compatible dimensions. The following assumption is required throughout the remainder of this paper. 
        \begin{assumption}\label{ass:AB2C2}
            $(A, B_2)$ is stabilizable and $(C_2, A)$ is detectable.
        \end{assumption}
  
         A full-order  Luenberger observer for system (\ref{LTIsys:stabilizing}) can be given by
		\begin{align}\label{observer}
			\dot{\hat x} &=A\hat x+B_2u+Lf,
		\end{align}
		where $f=C_2\hat x-y$ is the residual signal and $L$ is any matrix such that $A+LC_2$ is stable.
        Note that the Luenberger observer (\ref{observer}) is very related to the GIMC structure in Fig. \ref{fig:GIMC_disturbance}. Specifically, by letting $\bm{G}(s):=\left[\begin{array}{c|c}
			A & B_2\\
			\hline
			C_2 & 0
		\end{array}\right]$ in the GIMC structure, a state-space realization of $\bm{\tilde M}(s)$ and $\bm{\tilde N}(s)$ can be obtained as
		\begin{align*}
		\left[\begin{array}{cc}
			\bm{\tilde N}(s) & \bm{\tilde M}(s)
		\end{array}\right]=\left[\begin{array}{c|cc}
			A+LC_2 & B_2 & L\\
			\hline
			C_2 & 0 & I
		\end{array}\right],
		\end{align*}
		and it is not difficult to observe that $\bm{f}(s)=\bm{\tilde N}(s)\bm{u}(s)-\bm{\tilde M}(s)\bm{y}(s)$ is the residual signal generated by the Luenberger observer (\ref{observer}). 
  
		For the system $\bm G$ in (\ref{LTIsys:stabilizing}), consider a dynamic output-feedback stabilizing controller $\bm C$:
		\begin{align}\label{xc}
			{\bm C}:\left\{ \mspace{-6mu}\begin{array}{l}
					\dot x_c=A_cx_c+B_cy\\
					u_c=C_cx_c+D_cy.
			\end{array} \right.
		\end{align} 
 Clearly, $(A_c,B_c,C_c)$ is stabilizable and detectable, since $\bm C$ stabilizes $\bm G$ if and only if $\bm G$ stabilizes $\bm C$ \cite[Section 5.1]{vidyasagar2022control}. Then there exists a matrix $L_c$, such that $A_c+L_cC_c$ is stable. Motivated from the GIMC structure in Fig. \ref{fig:GIMC_disturbance}, we present the following composite controller $\bm {K_{CQ}}$:
		\begin{align}\label{KCQ}
			\bm {K_{CQ}}:\left\{ \mspace{-6mu}\begin{array}{l}
				\dot x_c =(A_c+L_cC_c)x_c-L_cu+(B_c+L_cD_c)y\\
				u_c=C_cx_c+D_cy\\
				\dot{\hat x} =A\hat x+B_2u+Lf\\
				u_q={\bm Q}(f)\\
				u=u_c+u_q
			\end{array} \right.
		\end{align}
		based on the observer (\ref{observer}) and the controller ${\bm C}$ in (\ref{xc}), where ${\bm Q}(\cdot)$ is a stable dynamic operator driven by the residual signal $f$ to be designed in the following form:
		\begin{align}\label{xq}
			{\bm Q}:\left\{ \mspace{-6mu}\begin{array}{l}
				\dot x_q=A_qx_q+B_qf\\
				u_q=C_qx_q+D_qf
			\end{array} \right.
		\end{align}
		with $A_q$ stable. 
	
\begin{figure}[!ht]
			\centering
			\includegraphics[scale=0.4]{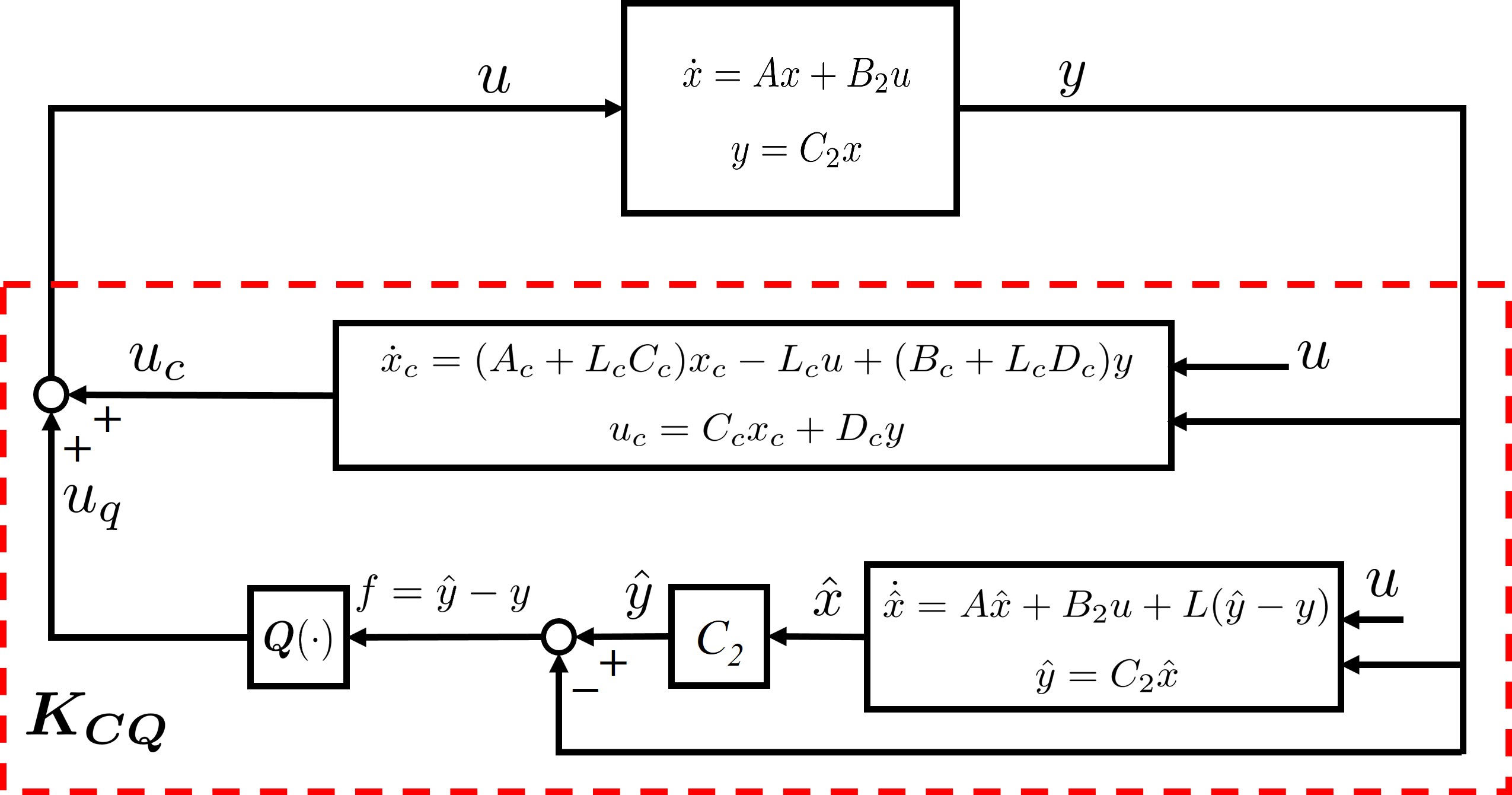}
			\caption{Two-controller structure.}
			\label{fig:KCQ_general}
		\end{figure}

		The significant feature of the composite controller $\bm {K_{CQ}}$ is that if the output signal $u_q$ of ${\bm Q}$ is zero, then $u=u_c$ and $\bm {K_{CQ}}$ is the same as the controller $\bm C$ in \eqref{xc}. 
       On the other hand, if $u_q\neq 0$, then the composite controller $\bm {K_{CQ}}$ behaves like a different one. Therefore, 
    $\bm {K_{CQ}}$ can be viewed as a two-controller structure to stabilize the system $\bm G$ in (\ref{LTIsys:stabilizing}), as shown in Fig. \ref{fig:KCQ_general}.

		Now consider another dynamic output-feedback stabilizing controller ${\bm K}$ designed in the following form:
		\begin{align}\label{xk}
			{\bm K}:\left\{ \mspace{-6mu}\begin{array}{l}
				\dot x_k=A_kx_k+B_ky\\
				u_k=C_kx_k+D_ky.
			\end{array} \right.
		\end{align} 
Then we give the solution of $\bm Q$ in state space such that 
$\bm{K_{CQ}}=\bm{K}$ when $u_q\ne 0$, which is presented in the following proposition. 
	
		\begin{proposition}\label{prop:generalQ}
		Given two arbitrarily designed stabilizing controllers $\bm{C}$ in (\ref{xc})  
and $\bm{K}$ in (\ref{xk})
and letting $L$ and $L_c$ be any matrices such that  both  $A+LC_2$ and $A_c+L_cC_c$ are stable, then a stable ${\bm Q}$ in (\ref{xq}) can be realized with
			\begin{align}\label{Qgeneral:ABCD}
                A_q&=\left[\begin{array}{ccc}
                    A_c+L_cC_c & (B_c+L_cD_q)C_2 & -L_cC_k \\
					0  &  A+B_2D_kC_2 & B_2C_k\\
				    0 & B_kC_2  & A_k
				\end{array}\right],\notag\\
				B_q&=\left[\begin{array}{ccc}
                    -B_c-L_cD_q \\
					L-B_2D_k\\
					-B_k
				\end{array}\right],\notag\\
				 C_q&=\left[\begin{array}{ccc}
				-C_c & -D_qC_2 &  C_k
				\end{array}\right],\notag\\
				D_q&=D_c-D_k,
			\end{align}
			such that $\bm {K_{CQ}}$ in (\ref{KCQ}) is stabilizing and $\bm{K_{CQ}}(s)=\bm{K}(s)$.
		\end{proposition}

        \begin{figure*}[!b]
			\normalsize
			\vspace*{4pt}
			\hrulefill
			\setcounter{MYtempeqncnt}{\value{equation}}
			\setcounter{equation}{7}
			\begin{align}\label{KCQ=K}
                \bm{K_{CQ}}(s)=\left[\begin{array}{ccccc|c}
					A+LC_2+B_2D_qC_2 & B_2C_c& 0 & -B_2D_qC_2 & B_2C_k & B_2D_k-L\\
					0 & A_c & 0 & -B_cC_2 & 0 & 0\\
                    0 & 0 & A_c+L_cC_c & (B_c+L_cD_q)C_2 & -L_cC_k & B_c+L_cD_q\\
                    0 & -B_2C_c & 0 & A+B_2D_cC_2 & 0 & 0\\
                    0 & 0 & 0 & B_kC_2 & A_k & B_k\\
					\hline
					0 & C_c & 0 & -D_qC_2 & C_k & D_k
				\end{array}\right]=\left[\begin{array}{c|c}
				A_k &  B_k\\
                \hline
				C_k & D_k
			\end{array}\right].
			\end{align}
			\setcounter{equation}{\value{MYtempeqncnt}}
		\end{figure*}
  
		\begin{IEEEproof}
            First, $A_q$ is obviously stable, as $A_c+L_cC_c$ and $\left[\begin{array}{cc}
				A+B_2D_kC_2 & B_2C_k   \\
				B_kC_2 & A_k 
			\end{array}\right]$ are both stable. Note that $\left[\begin{array}{cc}
				A+B_2D_kC_2 & B_2C_k   \\
				B_kC_2 & A_k 
			\end{array}\right]$ is the closed-loop matrix for the system $\bm G$ with controller $\bm K$. For the system $\bm G$ in (\ref{LTIsys:stabilizing}) with the controller $\bm {K_{CQ}}$ in (\ref{KCQ}), by letting $\left[\begin{array}{cccc}
			x'  & x_c' & x_q' & x'-\hat x'
		\end{array}\right]'$ as the closed-loop system state,  we obtain the following closed-loop matrix 
            \begin{align*}
            \left[\begin{array}{cccc}
				A+B_2D_cC_2 & B_2C_c & B_2C_q & - B_2D_qC_2  \\
				B_cC_2 & A_c & - L_cC_q & L_cD_qC_2  \\
				0   & 0 & A_q & B_qC_2 \\
                0  & 0 & 0 & A+LC_2\\
			\end{array}\right].
		\end{align*}
        The above matrix is stable, as $\left[\begin{array}{cc}
				A+B_2D_cC_2 & B_2C_c   \\
				B_cC_2 & A_c 
			\end{array}\right]$, $A_q$, and $A+LC_2$ are all stable. Thus $\bm {K_{CQ}}$ stabilizes $\bm G$. Next we shall show that if $\bm Q$ is realized by (\ref{Qgeneral:ABCD}), then $\bm{K_{CQ}}(s)=\bm{K}(s)$.
			It follows from the composite controller (\ref{KCQ}) that 
			\begin{align*}
				&\bm{K_{CQ}}(s)\notag\\
                =&\left[\begin{array}{ccc|c}
					A+LC_2+B_2D_qC_2 & B_2C_c&  B_2C_q & B_2D_k-L\\
					-L_cD_qC_2 & A_c & -L_cC_q & B_c+L_cD_q\\
                    B_qC_2& 0 & A_q & -B_q\\
					\hline
					D_qC_2 & C_c & C_q & D_k
				\end{array}\right],
			\end{align*}
			where the state is $\left[\begin{array}{ccc}
				\hat x' & x_c' & x_q' 
			\end{array}\right]'$. Substituting $(A_q, B_q, C_q, D_q)$ in (\ref{Qgeneral:ABCD}) into the above formula and using the similarity transformation 
			\begin{align*}
				X=\left[\begin{array}{ccccc}
					I & 0 & 0 & 0 & 0 \\
					0 & I & -I & 0 & 0 \\
					0 & 0 & I & 0  & 0\\
					-I & 0 & 0 & I & 0 \\
                    0 & 0 & 0 & 0 & I
				\end{array}\right],
			\end{align*}
			we obtain  $\bm{K_{CQ}}(s)=\left[\begin{array}{c|c}
				A_k &  B_k\\
                \hline
				C_k & D_k
			\end{array}\right]={\bm K}(s)$, where the detailed derivation is shown in \eqref{KCQ=K} in the bottom of the next page. 
		\end{IEEEproof}
		
		\addtocounter{equation}{1}

		\begin{remark}\label{rem:general}
			It follows from $(A_q, B_q, C_q, D_q)$ in (\ref{Qgeneral:ABCD}) that ${\bm Q}$ in fact consists of three parts in terms of controller $\bm{C}$,  the Luenberger observer (\ref{observer}), and controller $\bm{K}$.
			Specifically, writing $x_{ q}=\left[\begin{array}{ccc}
				x_{ q,1}' & x_{ q,2}' &  x_{ q,3}'
			\end{array}\right]'$, then ${\bm Q}$ can be described as
			\begin{align*}
                \dot x_{ q,1}=&(A_{ c}+L_cC_c)x_{ q,1}+(B_c+L_c(D_c-D_k))C_2x_{ q,2}\\
				&-L_cC_kx_{ q,3}-(B_{ c}+L_c(D_c-D_k))f,\\
				\dot x_{ q,2}=&(A+B_2D_kC_2)x_{ q,2}+B_2C_{ k}x_{ q,3}+(L-B_2D_k)f,\\
				\dot x_{ q,3}=&B_{ k}C_2x_{ q,2}+A_{ k}x_{ q,3}-B_{ k}f,\\
				u_{ q}=&(D_k-D_c)C_2x_{ q,2}-C_{ c}x_{ q,1}+C_kx_{ q,3}+(D_c-D_k)f.
			\end{align*}
			Considering $u=u_k$ in $\bm{K_{CQ}}$ described by \eqref{KCQ}, it is interesting to have the following relationship:
			\begin{align}\label{xq123}
				x_{ q,1}=x_{ c},\;\;  x_{ q,2}=\hat x,\ \ x_{ q,3}=x_k,
			\end{align}
			which suggests that the state of ${\bm Q}$ corresponds to the states of $\bm{C}$, the Luenberger observer (\ref{observer}), and $\bm{K}$. 
            Note that the parameters $(A_q, B_q, C_q, D_q)$ in \eqref{Qgeneral:ABCD} can be derived by letting $\bm{K_{CQ}}(s)=\bm{K}(s)$ and using the tool of linear fractional transformations \cite[Chapter 10]{zhou1996robust}.
            \pfbox
		\end{remark}

        \begin{remark}
            For the realization of $\bm Q$ in Proposition \ref{prop:generalQ}, if $A_c$ is stable, we can take $L_c=0$ in the controller $\bm{K_{CQ}}$ \eqref{KCQ} for simplicity and likewise we can take $L=0$ in the Luenberger observer \eqref{observer} if $A$ is stable. \pfbox 
        \end{remark}
        
		In the sequel, two interesting cases: shared observer and static stabilizing controllers, are considered. 
		
		\subsection{Two-controller structure with shared observer}
		
\begin{figure}[!ht]
			\centering
			\includegraphics[scale=0.4]{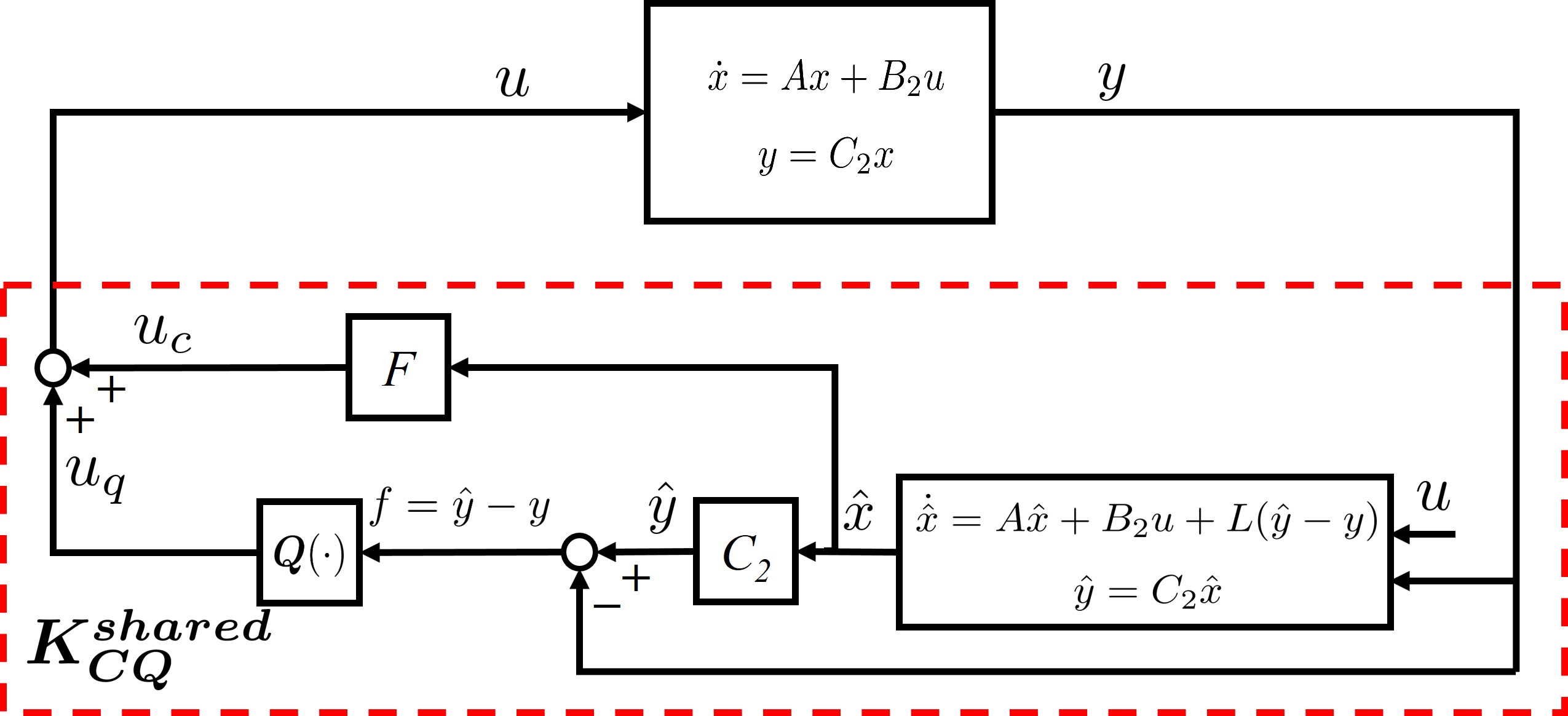}
			\caption{Two-controller structure with shared observer.}
			\label{fig:KCQ_shared}
		\end{figure}

		Now consider the controller ${\bm C}$ in (\ref{xc}) to be in the observer-based state-feedback structure with feedback gain $F$ and observer gain $L_o$, such that both $A+B_2F$ and $A+L_oC_2$ are stable. Then, with $A_c=A+B_2F+L_oC_2$, $B_c=-L_o$, $C_c=F$, $D_c=0$, and $L_c=-B_2$, the dynamic output-feedback controller (\ref{xc}) becomes the following observer-based state-feedback form:
		\begin{align}\label{xc:observer}
			{\bm C}: \left\{ \mspace{-6mu}\begin{array}{l}
				\dot x_c =(A+B_2F+L_oC_2)x_c-L_oy \\
				 u_c=Fx_c,
			\end{array} \right.
		\end{align}
		and the controller $\bm {K_{CQ}}$ in (\ref{KCQ}) is rewritten as
		\begin{eqnarray}
			\bm {K_{CQ}}:\left\{ \mspace{-6mu}\begin{array}{l}
				\dot x_c =Ax_c+B_2u+L_o(C_2x_c-y)\\
				u_c=F x_c\\
				\dot{\hat x} =A\hat x+B_2u+Lf\\
				u_q={\bm Q}(f)\\
				u=u_c+u_q.
			\end{array} \right.
		\end{eqnarray}
		Note that $A_c+L_cC_c=A+L_oC_2$ is stable here. A new controller structure with shared observer is proposed by forcing $L =L_o$ as follows:
		\begin{eqnarray}\label{KCQ:shared}
			\bm{K_{CQ}^{shared}}:\left\{ \mspace{-6mu}\begin{array}{l}
				\dot{\hat x} =A\hat x+B_2u+Lf\\
				u_c=F \hat x\\
				u_q={\bm Q}(f)\\
				u=u_c+u_q.
			\end{array} \right.
		\end{eqnarray}
	 The resulting control system is shown in Fig. \ref{fig:KCQ_shared}. 
		
		\begin{proposition}[Shared Observer]\label{prop:sharedQ}
		    Given an observer-based state-feedback stabilizing controller $\bm{C}$ in (\ref{xc:observer}) with $L_o =L$  and a separately and arbitrarily designed stabilizing controller $\bm{K}$ in (\ref{xk}), then a  stable ${\bm Q}$ in (\ref{xq}) can be realized with 
			\begin{align}\label{Qshared:ABCD}
				A_q&=\left[\begin{array}{cc}
					A+B_2D_kC_2 & B_2C_{ k} \\ 
					B_{ k}C_2 & A_{ k}
				\end{array}\right],\; B_{q}=\left[\begin{array}{c}
					L-B_2D_k\\ 
					-B_k
				\end{array}\right],\notag\\
				C_q&=\left[\begin{array}{cc}
					D_kC_2 -F & C_{ k}
				\end{array}\right],\; D_q=-D_k,
			\end{align}
			such that $\bm{K_{CQ}^{shared}}$ in (\ref{KCQ:shared}) is stabilizing and $\bm{K_{CQ}^{shared}}(s)=\bm{K}(s)$. 
		\end{proposition}
		
		\begin{IEEEproof}
			The proof is similar to that of Proposition \ref{prop:generalQ} and thus is omitted. In fact, the formulas of $(A_q, B_q, C_q, D_q)$ can be obtained directly by substituting $A_c=A+B_2F+LC_2$, $B_c=-L$, $C_c=F$, $D_c=0$, and $L_c=-B_2$ into (\ref{Qgeneral:ABCD}), after removing the (stable) uncontrollable mode. 
		\end{IEEEproof}

		\begin{remark}
			As in Remark \ref{rem:general}, ${\bm Q}$ in (\ref{Qshared:ABCD}) can be interpreted as follows.
			Write $x_{ q}=\left[\begin{array}{cc}
				x_{ q,1}' & x_{ q,2}' 
			\end{array}\right]'$, such that $\bm Q$ can be described as
			\begin{align*}
				\dot x_{ q,1}&=(A+B_2D_kC_2)x_{ q,1}+B_2C_{ k}x_{ q,2}+(L-B_2D_k)f,\\
				\dot x_{ q,2}&=B_{ k}C_2x_{ q,1}+A_{ k}x_{ q,2}-B_{ k}f,\\
				u_q&=(D_kC_2 -F)x_{ q,1} + C_kx_{ q,2}-D_kf.
			\end{align*}
			Likewise, considering $u=u_k$ in $\bm{K_{CQ}^{shared}}$ described by \eqref{KCQ:shared}, it can be easily checked that 
			\begin{align}\label{xq123:observer}
				x_{ q,1}=\hat x,\ \ x_{ q,2}=x_{ k},
			\end{align}
			suggesting that the state of ${\bm Q}$ corresponds to the states of the Luenberger observer (\ref{observer}) and the controller $\bm{K}$.  \pfbox
		\end{remark}

		\subsection{Two-controller structure with static controllers}\label{subsec:static}
\begin{figure}[!ht]
			\centering
			\includegraphics[scale=0.4]{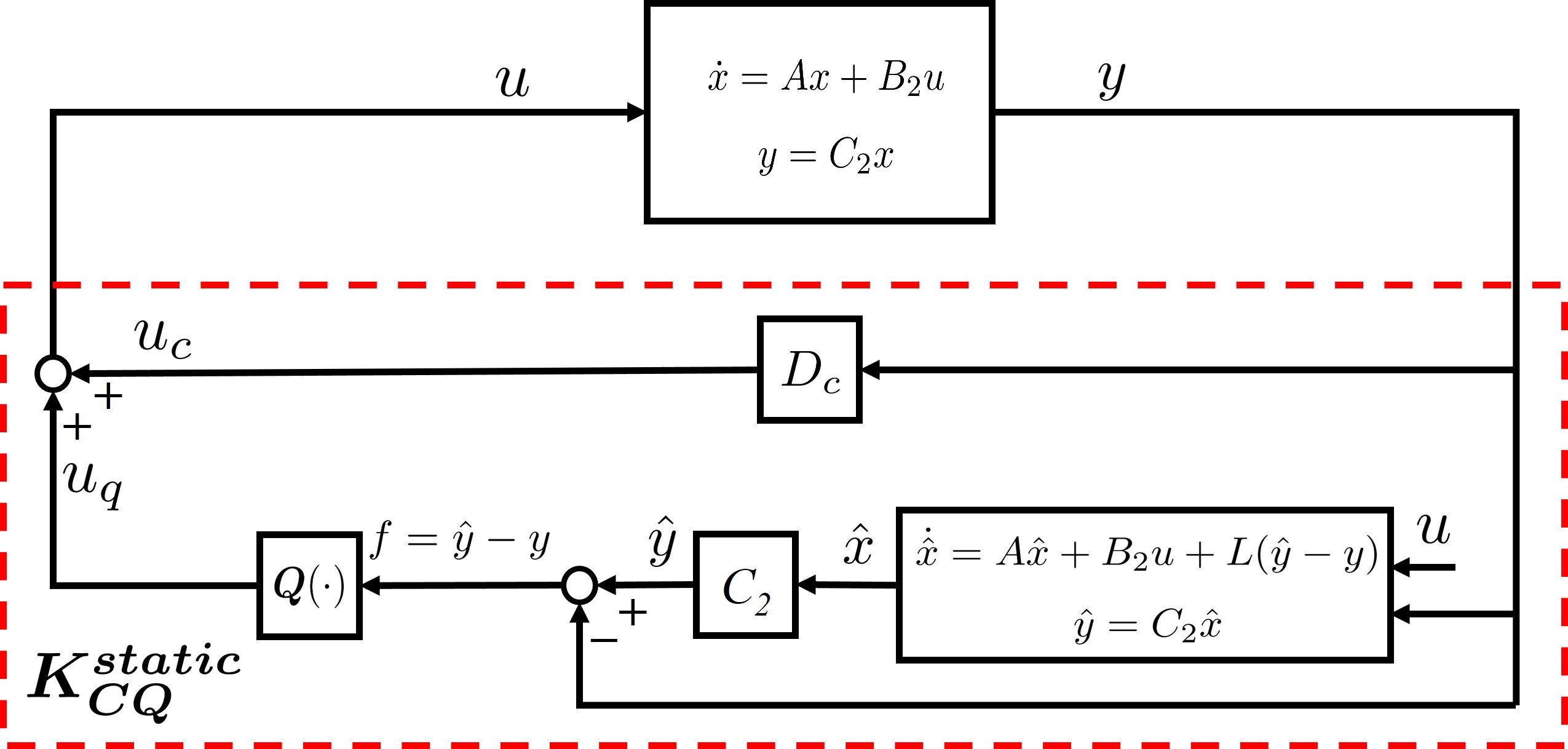}
			\caption{Two-controller structure with static controllers.}
			\label{fig:KCQ_static}
		\end{figure}

		Assume that there exists a static controller stabilizing system (\ref{LTIsys:stabilizing}). Then let controller ${\bm C}$ be static: 
		\begin{align}\label{xc:static}
			{\bm C:} u_c=D_{c}y.
		\end{align}
  The controller $\bm {K_{CQ}}$ in (\ref{KCQ}) becomes
		\begin{eqnarray}\label{KCQ:static}
			\bm{K_{CQ}^{static}}:\left\{ \mspace{-6mu}\begin{array}{l}
                u_c=D_c y\\
				\dot{\hat x} =A\hat x+B_2u+Lf\\
				u_q={\bm Q}(f)\\
				u=u_c+u_q.
			\end{array} \right.
		\end{eqnarray}
		The resulting control system  is shown in Fig. \ref{fig:KCQ_static}. The solution to ${\bm Q}$ in state space is presented in the following proposition and its proof is similar to that of the previous two propositions and thus is omitted.
        \begin{proposition}[Static Controller]\label{prop:static}
		    Given a static stabilizing controller $\bm{C}$ in (\ref{xc:static})  and a separately and arbitrarily designed stabilizing controller $\bm{K}$ in (\ref{xk}), then a  stable ${\bm Q}$ in (\ref{xq}) can be realized with 
		\begin{align}
			A_q&=\left[\begin{array}{cc}
				A+B_2D_kC_2 & B_2C_{ k} \\ 
				B_{ k}C_2 & A_{ k}
			\end{array}\right],\; B_{q}=\left[\begin{array}{c}
				L-B_2D_k\\ 
				-B_k
			\end{array}\right],\notag\\
			C_q&=\left[\begin{array}{cc}
				(D_k-D_c)C_2 & C_{ k}
			\end{array}\right],\; D_q=D_c-D_k.
		\end{align}
			such that $\bm{K_{CQ}^{static}}$ in (\ref{KCQ:static}) is stabilizing and $\bm{K_{CQ}^{static}}(s)=\bm{K}(s)$. 
		\end{proposition}

  If controller ${\bm K}$ is also static, i.e., $u_k=D_ky$, then ${\bm Q}$ reduces to
		\begin{align}\label{Q:sf}
			A_q&=A+B_2D_kC_2,\; B_{q}=L-B_2D_k,\notag\\
			C_q&=(D_k-D_c)C_2,\; D_q=D_c-D_k.
		\end{align} 
		
		\emph{State feedback:} The static state feedback is a special static case when $C_2=I$. In this case, the observer gain $L$ can be simply chosen as $L=B_2D_c$, as $A+LC_2=A+B_2D_c$ is stable.
		
		\section{Multi-Objective Complementary Control (MOCC)}\label{sec:MOCC}
		In general, a typical MO control problem specifies different objectives on different channels of the closed-loop
		system \cite{bernstein1989lqg,zhou1994mixed,doyle1994mixed,scherer1997multiobjective,hindi1998multiobjective,chen2001multiobjective,balandin2019multi}. The objectives under consideration can be roughly divided into two classes: performance involving commands and robustness against unknown disturbances. Performance is meant for transient process, tracking accuracy, passivity requirement \cite{keemink2018admittance}, etc, while robustness is meant to keep healthy performance in an uncertain or partially unknown environment. 
  It has been pointed out in the Introduction that the traditional design technique using a single controller to address multiple objectives of the closed-loop system usually renders trade-off solutions with compromised performance and robustness in some sense, and for 2DoF controllers there does not exist a general and systematic design method so far to address both performance and the robustness.  In contrast, we shall show that MO
control problems can be handled effectively by applying the
two-controller structures presented in the previous
section with two independently designed controllers operating
in a naturally complementary way, leading to a MO complementary control (MOCC) framework. 
  
        In the sequel,  a specific robust tracking control problem is presented  to demonstrate the advantages of the proposed MOCC framework.
The MO tracking problem under consideration has two objectives: tracking performance and robustness. In the new control structure, $\bm C$ is designed to address the tracking performance without disturbances, and $\bm K$ is designed to address the robustness with respect to unknown/uncertain disturbances. 

Consider a perturbed FDLTI system (\ref{LTIsys:stabilizing}), described by
		\begin{align}\label{LTIsys:w}
            \left\{ \mspace{-6mu}\begin{array}{ll}
					\dot x\mspace{-12mu}&=Ax+B_1w+B_2u\\
                z\mspace{-12mu}&=C_1(C_2x-r)+D_{12}u\\
    			y\mspace{-12mu}&=C_2x+D_{21}w,
			\end{array} \right.
		\end{align}
		where  $x\in\mathbb{R}^n$ is the system state, $u\in\mathbb{R}^{m_2}$ is the control input, $y\in\mathbb{R}^{p_2}$ is the measured output, $w\in\mathbb{R}^{m_1}$ is the unknown disturbance input, $r\in\mathbb{R}^{p_2}$ known or measurable reference signal, and $z\in\mathbb{R}^{p_1}$ consisting of tracking error and control input is an output variable evaluating the tracking performance.

\begin{figure}[!ht]
			\centering
			\includegraphics[scale=0.4]{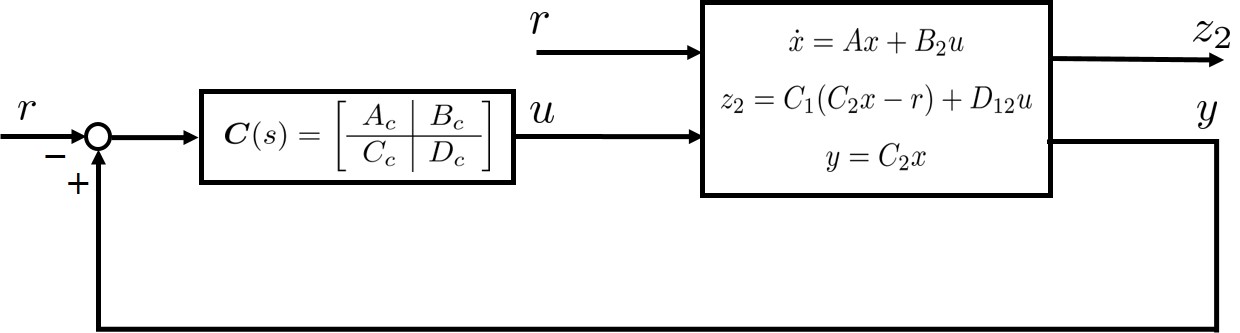}
			\caption{Design of the nominal tracking controller ${\bm C}$.}
			\label{fig:C_tracking}
		\end{figure}

\begin{figure}[!ht]
			\centering
			\includegraphics[scale=0.4]{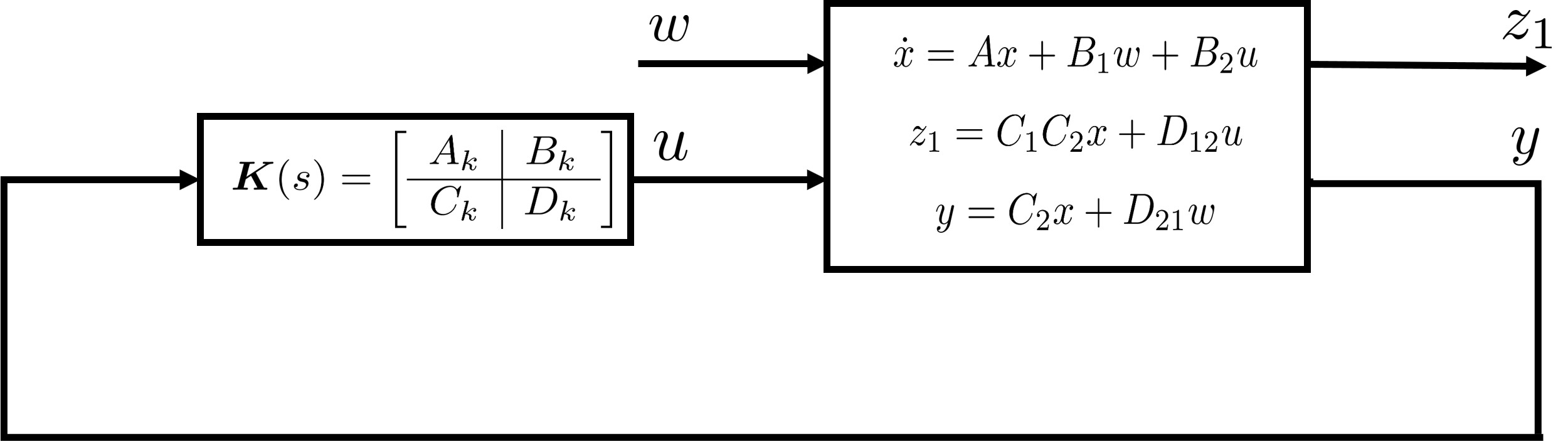}
			\caption{Design of the robust controller ${\bm K}$.}
			\label{fig:K_robust}
		\end{figure}

\begin{figure}[!ht]
			\centering
			\includegraphics[scale=0.4]{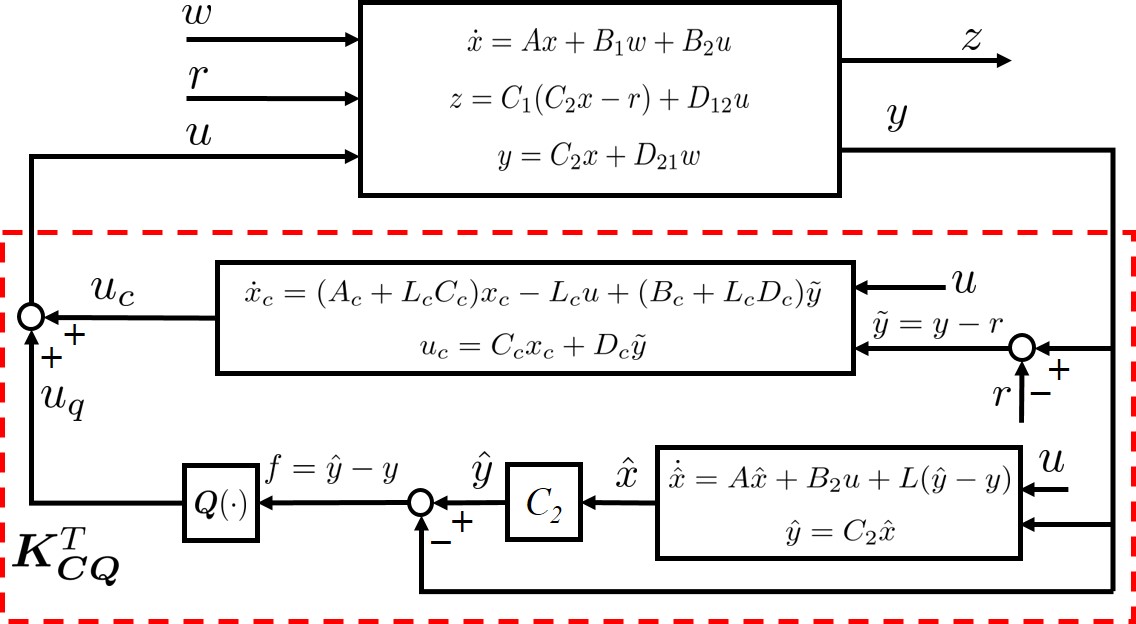}
			\caption{MO tracking control system by two-controller structure.}
			\label{fig:KCQ_tracking}
		\end{figure}

Assume that an admissible tracking controller ${\bm C}$ is designed as the following form 
    \begin{align}\label{C:tracking}
			{\bm C}:\left\{ \mspace{-6mu}\begin{array}{l}
					\dot x_c=A_cx_c+B_c(y-r)\\
					u_c=C_cx_c+D_c(y-r),
			\end{array} \right.
		\end{align}
	for the system in (\ref{LTIsys:w}) with $w=0$, i.e.,
    \begin{align}\label{LTIsys:w0}
        \left\{ \mspace{-6mu}\begin{array}{ll}
					\dot x\mspace{-12mu}&=Ax+B_2u\\
                z_2\mspace{-12mu}&=C_1(C_2x-r)+D_{12}u\\
    			y\mspace{-12mu}&=C_2x,
			\end{array} \right.
		\end{align}
 as shown in Fig. \ref{fig:C_tracking}.
  On the other hand, assume that a robust controller ${\bm K}$ having the form of
    \begin{align}\label{K:tracking}
			{\bm K}:\left\{ \mspace{-6mu}\begin{array}{l}
					\dot x_k=A_kx_k+B_ky\\
					u_k=C_kx_k+D_ky
			\end{array} \right.
		\end{align}
   is designed for the system in (\ref{LTIsys:w}) with $r=0$, i.e,
   \begin{align}\label{LTIsys:r0}
            \left\{ \mspace{-6mu}\begin{array}{ll}
					\dot x\mspace{-12mu}&=Ax+B_1w+B_2u\\
                z_1\mspace{-12mu}&=C_1C_2x+D_{12}u\\
    			y\mspace{-12mu}&=C_2x+D_{21}w,
			\end{array} \right.
		\end{align}
   as shown in Fig. \ref{fig:K_robust}. 
         Then by applying the two-controller structure in Fig. \ref{fig:KCQ_general}, a MO tracking controller $\bm{K_{CQ}^{T}}$ (the superscript ``$T$'' stands for tracking) can be obtained shown in Fig. \ref{fig:KCQ_tracking}, where the state-space model of $\bm{Q}$ is constructed as that in Proposition \ref{prop:generalQ}:  
         \begin{align*}
			{\bm Q}:\left\{ \mspace{-6mu}\begin{array}{l}
				\dot x_q=A_qx_q+B_qf\\
				u_q=C_qx_q+D_qf,
			\end{array} \right.
		\end{align*}
        with $(A_q, B_q, C_q, D_q)$ given by (\ref{Qgeneral:ABCD}). Note that the difference between the controller $\bm {K_{CQ}}$ in Fig. \ref{fig:KCQ_general} and the tracking controller $\bm {K_{CQ}^T}$ in Fig. \ref{fig:KCQ_tracking} is that for $\bm {K_{CQ}^T}$, the reference signal $r$ is also as an input to generate the control signal $u_c$. Then it follows from Proposition \ref{prop:generalQ} that for the tracking control system with $\bm{K_{CQ}^{T}}$, the transfer matrix from $y$ to $u$ is $\bm K(s)$.
    
        \begin{remark}
            As the two-controller structure studied in Section \ref{sec:tcs}, the tracking controller $\bm {K_{CQ}^T}$ can also have shared-observer and static structures. If the tracking controller $\bm C$ in \eqref{C:tracking} is observer-based with  $A_c=A+B_2F+LC_2$, $B_c=-L$, $C_c=F$, $D_c=0$, and $L_c=-B_2$, then $u_c$ in $\bm {K_{CQ}^T}$ becomes
            \begin{align*}
            u_c=Fx_c, \; \dot x_c =Ax_c+B_2u+L(C_2x_c-y+r).
             \end{align*}
            Write $x_c$ above as a sum of two terms: $x_c=\tilde{x}_c+x_r$, such that   
        \begin{align*}
            \dot {\tilde x}_c &=A\tilde x_c+B_2u+L(C_2\tilde x_c-y),\\
            \dot x_r &=(A+LC_2)x_r+Lr.
        \end{align*}
        Then similar to the controller $\bm{K_{CQ}^{shared}}$ in \eqref{KCQ:shared}, the tracking controller $\bm {K_{CQ}^T}$ with shared observer is described as 
        \begin{eqnarray*}
			\bm{K_{CQ}^{shared,T}}: \left\{ \mspace{-6mu}\begin{array}{l}
				\dot{\hat x} =A\hat x+B_2u+Lf\\
                \dot x_r =(A+LC_2)x_r+Lr\\
				u_c=F (\hat x+x_r)\\
				u_q={\bm Q}(f)\\
				u=u_c+u_q.
			\end{array} \right.
		\end{eqnarray*}
        The parameters $(A_q, B_q, C_q, D_q)$ of $\bm Q$ are given by Proposition \ref{prop:sharedQ}. If the tracking controller $\bm C$ in \eqref{C:tracking} is static as $u_c=D_c(y-r)$, the tracking controller $\bm {K_{CQ}^T}$ is simply given by
        \begin{eqnarray*}
			\bm{K_{CQ}^{static,T}}:\left\{ \mspace{-6mu}\begin{array}{l}
                u_c=D_c(y-r)\\
				\dot{\hat x} =A\hat x+B_2u+Lf\\
				u_q={\bm Q}(f)\\
				u=u_c+u_q
			\end{array} \right.
		\end{eqnarray*}
        with $(A_q, B_q, C_q, D_q)$ given by Proposition \ref{prop:static}.
        \pfbox
        \end{remark}
  
Next, we shall conduct a performance analysis for the resulting tracking controller $\bm{K_{CQ}^{T}}$.  It will be shown that the tracking controller $\bm{K_{CQ}^{T}}$ consisting of ${\bm C}$ for the nominal tracking performance and ${\bm K}$ for the robustness leads to a decoupled design and the two controllers ${\bm C}$ and ${\bm K}$ operate in a complementary way.
		
		\subsection{Performance analysis}\label{subsec:analysis}
		In principle, the objective of performance analysis in this paper is to determine the closed-loop performance generated by $\bm{K_{CQ}^{T}}$ under certain specified criteria such as the $L_1$ norm and $L_2$ norm of $z$. Here, we shall choose the power norm of bounded power signals, since it can be used for persistent signals and is related to the quadratic criterion and the $\mathcal{H}_{\infty}$ norm.  It is common to use the power norm of bounded power signals in the analysis of control systems, e.g. \cite{zhou1994mixed,doyle1994mixed,chen2001multiobjective,makila1998bounded,wan2013integrated}. All signals considered in this paper are assumed to be deterministic. In the sequel, the notion of bounded power signals is introduced.
  
        Given a real vector signal $u(t)$ that is zero for $t<0$, its asymptotically stationary autocorrelation matrix is defined as 
		\begin{align*}
			R_{uu}(\tau)=\lim\limits_{T\to\infty}\frac{1}{T}\int_{0}^{T} u(t+\tau)u(t)' {\rm d}t.
		\end{align*} 
		The Fourier transform of $R_{uu}(\tau)$ called the (power) spectral density of $u$, if exists, is  
		\begin{align*}
			\bm{S_{uu}}(j\omega):=\int_{-\infty}^{\infty}R_{uu}(\tau)e^{-j\omega\tau}{\rm d}\tau.
		\end{align*}
		Then the bounded power signals are defined as follows. 
		
		\begin{definition}[\cite{zhou1994mixed}]
			A signal $u$ is said to have bounded power if it satisfies the following conditions:
			\begin{itemize}
				\item[1.] $\|u(t)\|$ is bounded for all $t\ge 0$;
				\item[2.] The autocorrelation matrix $R_{uu}(\tau)$ exists for all $\tau$ and the spectral density matrix $\bm{S_{uu}}(j\omega)$ exists;
				\item[3.] $\lim\limits_{T\to\infty}\frac{1}{T}\int_{0}^{T} \|u(t)\|^2 {\rm d}t <\infty$.
			\end{itemize}
		\end{definition}
		
		The set of all signals having bounded power is denoted by $\mathcal{P}$. A seminorm can be defined on $\mathcal{P}$:
		\begin{align}
			\|u\|_{\mathcal P}&=\sqrt{\lim\limits_{T\to\infty}\frac{1}{T}\int_{0}^{T} \|u(t)\|^2 {\rm d}t}\notag\\
			&=\sqrt{{\rm Trace}[R_{uu}(0)]} \;\; \forall u\in\mathcal{P}.
		\end{align}
		The power seminorm of a signal can also be computed from its spectral density matrix by using the inverse Fourier transform of $R_{uu}(\tau)$:
		\begin{align}\label{up:frequency}
			\|u\|_{\mathcal P}=\sqrt{\frac{1}{2\pi}\int_{-\infty}^{\infty}{\rm Trace}[\bm{S_{uu}}(j\omega)]{\rm d}\omega}=\|\bm u(s)\|_{\cal P},
		\end{align}
        which is derived from the Parseval theorem in the average power case \cite[Section 3.5.7]{oppenheim1997signals} with $\bm u(s)$ being the Laplace transform of $u(t)$.
		The cross-correlation between two real signals $u$ and $v$ is defined as
		\begin{align*}
			R_{uv}(\tau)=\lim\limits_{T\to\infty}\frac{1}{T}\int_{0}^{T} u(t+\tau)v(t)' {\rm d}t,
		\end{align*}
		and its Fourier transform is denoted by $\bm{S_{uv}}(j\omega)$, called cross (power) spectral density.
		
		Now consider the tracking control system in Fig. \ref{fig:KCQ_tracking}, with $\bm Q$ constructed according to Proposition \ref{prop:generalQ}. Both the disturbance signal $w$ and the reference signal $r$ are assumed to be bounded power signals, i.e.,  $w, r\in\mathcal{P}$. The closed-loop performance will be measured in terms of the power norm of the performance output $z$, i.e., $\|z\|_\mathcal{P}$. 
        First, the closed-loop system can be described by
		\begin{align}\label{sys:cl}
			\bm{T}:\left\{ \mspace{-6mu}\begin{array}{l}
				\dot{\bar x} =\bar A\bar x+\bar B_1w+\bar B_rr\\
				z=\bar C_1\bar x+D_rr,
			\end{array} \right.
		\end{align}
		where ${\bar x} :=\left[\begin{array}{cccc}
			x' & x'-\hat x' & x_c' & x_q'
		\end{array}\right]'$ and
		\begin{align*}
			\bar A&=\left[\begin{array}{cccc}
				A+B_2D_cC_2 & -B_2D_qC_2& B_2C_c & B_2C_q \\
				0 & A+LC_2 & 0 & 0 \\
				B_cC_2 & L_cD_qC_2 & A_c & -L_cC_q\\
				0 & -B_qC_2 & 0 & A_q\\
			\end{array}\right],\\ \bar B_1&=\left[\begin{array}{c}
				B_1+B_2(D_c-D_q)D_{21}\\
				B_1+LD_{21}\\
				(B_c+L_cD_q)D_{21}\\
				-B_qD_{21}
            \end{array}\right],\; \bar B_r=\left[\begin{array}{c}
				-B_2D_c\\
				0\\
				-B_c\\
				0
            \end{array}\right],\\
			\bar C_1&=\left[\mspace{-6mu}\begin{array}{ccccc}
				(C_1+D_{12}D_c)C_2 & -D_{12}D_qC_2 & D_{12}C_c & D_{12}C_q
			\end{array}\mspace{-6mu}\right],\\
			D_r&=-(C_1+D_{12}D_c).
		\end{align*}
        The following decoupled result is important in this paper, which characterizes the nature of the closed-loop system with the controller $\bm{K_{CQ}^{T}}$.

        \begin{figure*}[!b]
			\normalsize
			\vspace*{4pt}
			\hrulefill
			\setcounter{MYtempeqncnt}{\value{equation}}
			\setcounter{equation}{27}
			\begin{align}\label{Ts:proof}
                \bm{T}(s)&=\left[\begin{array}{c|cc}
					\mathcal A & \mathcal B_1 & \mathcal B_r\\
					\hline
					\mathcal C_1& D_1 & D_r
				\end{array}\right]\notag\\
				=&\left[\begin{array}{cccccc|cc}
					A+B_2D_cC_2 & 0  & B_2C_c  &  0 & 0 & 0 &  0 & -B_2D_c\\
					0 & A+LC_2  & 0 & 0 & 0 & 0 & B_1+LD_{21} & 0 \\
					B_cC_2 & 0 & A_c & 0 & 0 & 0 & 0 & -B_c\\
                    0 & 0 & 0  & A_c+L_cC_c & (B_c+L_cD_q)C_2 & -L_cC_k & (B_c+L_cD_q)D_{21} & 0\\
                    0 & 0 & 0  & 0 & A+B_2D_kC_2 & B_2C_k & B_1+B_2D_kD_{21} & 0\\
                    0 & 0 & 0 & 0 & B_kC_2  & A_k & B_kD_{21} & 0\\
					\hline
					(C_1+D_{12}D_c)C_2 & 0 & D_{12}C_c & 0 & (C_1+D_{12}D_k)C_2  & D_{12}C_k & 0 & D_r
                    \end{array}\right].
			\end{align}
			\setcounter{equation}{\value{MYtempeqncnt}}
		\end{figure*}
        
        \begin{lemma}\label{lem:cl}
            The closed-loop system (\ref{sys:cl}) can be represented as the following form of transfer matrix in terms of state-space matrices:
            \begin{align}
			\bm{T}(s)&=\left[\begin{array}{c|cc}\bar A & \bar B_1 & \bar B_r \\
				\hline
				\bar C_1& 0 & D_r
			\end{array}\right]=\left[\begin{array}{cc}
				\bm{T_{z_1w}}(s) & \bm{T_{z_2r}}(s)
			\end{array}\right],
		\end{align}
            where
            \begin{align*}
				\bm{T_{z_1w}}(s)\mspace{-6mu}&=\mspace{-6mu}\left[\mspace{-6mu}\begin{array}{cc|c}
					A+B_2D_kC_2 & B_2C_k  & B_1+B_2D_kD_{21} \\
					B_kC_2 & A_k & B_kD_{21}\\
					\hline
					(C_1+D_{12}D_k)C_2 & D_{12}C_k & 0
				\end{array}\mspace{-6mu}\right]\\
                \bm{T_{z_2r}}(s)\mspace{-6mu}&=\mspace{-6mu}\left[\mspace{-6mu}\begin{array}{cc|c}
					A+B_2D_cC_2 & B_2C_c  & -B_2D_c \\
					B_cC_2 & A_c & -B_c\\
					\hline
					(C_1+D_{12}D_c)C_2 & D_{12}C_c & -(C_1+D_{12}D_c)
				\end{array}\mspace{-6mu}\right],
			\end{align*}
        \end{lemma}
         with $z_1$ and $z_2$ defined in (\ref{LTIsys:r0}) and (\ref{LTIsys:w0}), respectively.

        \begin{IEEEproof}
            Substituting the formulas of $A_q, B_q, C_q, D_q $ in (\ref{Qgeneral:ABCD}) into the closed-loop system (\ref{sys:cl}) and using the linear transformation $\chi=X\bar x$ with 
			\begin{align*}
				X=\left[\begin{array}{cccccc}
					I & -I & 0 & 0 & -I & 0\\
					0 & I & 0 & 0 & 0 & 0\\
					0 & 0 & I & -I &  0 & 0\\
					0 & 0 & 0 & I & 0 & 0\\
                    0 & I & 0 & 0 & I & 0\\
					0 & 0 & 0 & 0 & 0 & I
				\end{array}\right],
			\end{align*}
			we have 
			\begin{align*}
				\bm{T}(s)&=\left[\begin{array}{c|cc}
					\mathcal A & \mathcal B_1 & \mathcal B_r\\
					\hline
					\mathcal C_1& 0 & D_r
				\end{array}\right],
			\end{align*}
			where  $\mathcal A=X\bar AX^{-1}, \mathcal B_1=X\bar B_1, \mathcal B_r=X\bar B_r, \mathcal C_1=\bar C_1X^{-1}$ and their detailed expressions are shown in \eqref{Ts:proof} in the bottom of the next page. Then, it can be readily obtained the state-space data of $\bm{T_{z_1w}}(s)$ and $\bm{T_{z_2r}}(s)$ in the theorem.
        \end{IEEEproof}
        \addtocounter{equation}{1} 
        
		It is clearly seen that the closed-loop transfer function ${\bm T}(s)$ consists of $\bm{T_{z_1w}}(s)$ and $\bm{T_{z_2r}}(s)$ which depend on the controller ${\bm K}$ and the controller ${\bm C}$ separately, and $\bm T(s)$ does not depend on $L$ and $L_c$. Note that $\bm{T_{z_1w}}(s)$ and $\bm{T_{z_2r}}(s)$ are the closed-loop transfer matrices in Figs. \ref{fig:K_robust} and \ref{fig:C_tracking}, respectively.  The total performance achieved by $\bm K$ and $\bm C$ is characterized in the analysis below, through the power norm of $z$. Due to the neat separate transfer matrices $\bm{T_{z_1w}}(s)$ and $\bm{T_{z_2r}}(s)$, we shall analyze $\|z\|_{\mathcal P}$ in the frequency domain.
  
    Defining $v:=\left[\begin{array}{cc}
			w\\ r
		\end{array}\right]$, then the spectral density matrix of $v$ can be written as
		\begin{align*}
			\bm{S_{vv}}(j\omega) =\left[\begin{array}{cc}
				\bm{S_{ww}}(j\omega)  & \bm{S_{wr}}(j\omega) \\
				\bm{S_{wr}^*}(j\omega)  & \bm{S_{rr}}(j\omega) 
			\end{array}\right].
		\end{align*}
		From the spectral analysis \cite{zhou1994mixed} and (\ref{up:frequency}), we get
		\begin{align}\label{Szz}
			\bm{S_{zz}}=\left[\begin{array}{cc}
				\bm{T_{z_1w}} & \bm{T_{z_2r}}
			\end{array}\right]\left[\begin{array}{cc}
				\bm{S_{ww}} & \bm{S_{wr}}\\
				\bm{S_{wr}^*} & \bm{S_{rr}}
			\end{array}\right]\left[\begin{array}{c}
				\bm{T_{z_1w}^*} \\ \bm{T_{z_2r}^*}
			\end{array}\right]
		\end{align} 
        and
        \begin{align}\label{zp:omega}
			\|z\|_{\mathcal P}=\sqrt{\frac{1}{2\pi}\int_{-\infty}^{\infty}{\rm Trace}[\bm{S_{zz}}(j\omega)]{\rm d}\omega}.
		\end{align}
        Generally, the disturbance signal $w$ and the reference signal $r$ could be independent or dependent on each other. Signals $w$ and $r$ are said to be orthogonal or independent if $\bm{S_{wr}}=0$ \cite{gardner1988statistical,zhou1994mixed}. For example, any two sinusoidal signals $u$ and $v$ with different frequencies are orthogonal since the cross-correlation matrix $R_{uv}(\tau)=0, \forall \tau$ which is the orthogonal case. The dependent case could happen when $w$ is induced by plant parameter uncertainties, which in turn is related to the reference signal $r$ in some way. Performance evaluations are carried out for both orthogonal and dependent cases and summarized in the following two theorems.  
       
		\begin{theorem}\label{thm:orthogonal}
			Let $w$ and $r$ be orthogonal, that is, $\bm{S_{wr}}=0$. Given any controllers $\bm C$ and $\bm K$ in (\ref{C:tracking}) and (\ref{K:tracking}), then the performance measure $\|z\|_{\mathcal{P}}^2$ for the closed-loop system (\ref{sys:cl}) is sum of two separate terms:
			\begin{align}
	           \|z\|_{\mathcal{P}}^2&=\|z_1\|_{\mathcal{P}}^2+\|z_2\|_{\mathcal{P}}^2,
			\end{align} 
			where 
            \begin{align*}
           \|z_1\|_{\mathcal{P}}^2&=\frac{1}{2\pi}\int_{-\infty}^{\infty}{\rm Trace}[\bm{T_{z_1w}}(j\omega)\bm{S_{ww}}(j\omega)\bm{T_{z_1w}}^*(j\omega)]{\rm d}\omega, \\
            \|z_2\|_{\mathcal{P}}^2&=\frac{1}{2\pi}\int_{-\infty}^{\infty}{\rm Trace}[\bm{T_{z_2r}}(j\omega)\bm{S_{rr}}(j\omega)\bm{T_{z_2r}^*}(j\omega)]{\rm d}\omega,
            \end{align*}
        with $\bm{T_{z_1w}}(s)$ and $\bm{T_{z_2r}}(s)$ dependent on ${\bm K}$ and  ${\bm C}$ respectively according to Lemma \ref{lem:cl}.
		\end{theorem}
  
		\begin{IEEEproof}
			The proof can be done by direct calculation. Since $\bm{z}=\bm {T_{z_1w_1}}\bm{w}+\bm {T_{z_2r}}\bm{r}$, if $w$ and $r$ are orthogonal, i.e., $\bm{S_{wr}}=0$, recalling $\bm{S_{zz}}(j\omega)$ and  $\|z\|_{\mathcal{P}}$ given in (\ref{Szz}) and (\ref{zp:omega}), we immediately have
			\begin{align*}
				\|z\|_\mathcal{P}^2=&\frac{1}{2\pi}\int_{-\infty}^{\infty}{\rm Trace}[\bm{S_{zz}}(j\omega)]{\rm d}\omega\\
				=& \frac{1}{2\pi}\int_{-\infty}^{\infty}\Big({\rm Trace}[\bm{T_{z_1w}}(j\omega)\bm{S_{ww}}(j\omega)\bm{T_{z_1w}^*}(j\omega)]\\
				&+{\rm Trace}[\bm{T_{z_2r}}(j\omega)\bm{S_{rr}}(j\omega)\bm{T_{z_2r}^*}(j\omega)]\Big){\rm d}\omega,\\
                =&\|z_1\|_\mathcal{P}^2+\|z_2\|_\mathcal{P}^2,
			\end{align*}
             where $\|z_2\|_\mathcal{P}^2=\frac{1}{2\pi}\int_{-\infty}^{\infty}{\rm Trace}[\bm{T_{z_2r}}(j\omega)\bm{S_{rr}}(j\omega)\bm{T_{z_2r}^*}(j\omega)]{\rm d}\omega$ and $\|z_1\|_{\mathcal{P}}^2=\frac{1}{2\pi}\int_{-\infty}^{\infty}{\rm Trace}[\bm{T_{z_1w}}(j\omega)\bm{S_{ww}}(j\omega)\bm{T_{z_1w}}^*(j\omega)]{\rm d}\omega$. 
		\end{IEEEproof}
        
	\begin{remark}\label{rem:performance_orthogonal}
    It is clear that the nominal tracking performance is characterized in $\|z_2\|_{\cal P}$ with respect to the reference signal $r$ and the impact of disturbance $w$ (hence, the robustness) is measured in $\|z_1\|_{\cal P}$. Since $\bm{T_{z_1w}}$ only depends on $\bm K$ and $\bm{T_{z_2r}}$ is only related to $\bm C$, Theorem \ref{thm:orthogonal} shows that if $w$ and $r$ are orthogonal, there is no trade-off between the design of ${\bm C}$ for the nominal tracking performance $\|z_2\|_{\cal P}$ and the design of ${\bm K}$ for the robustness.
     In other words, the two independently designed controllers ${\bm C}$ and ${\bm K}$ operate in a naturally complementary way, and, 
     obviously, when $w=0$, the nominal tracking performance can be fully kept by $\bm C$. Furthermore, it can also be seen that, to achieve better robust tracking performance, $\bm C$ should be designed to bring $\|z_2\|_{\cal P}$ low, while $\bm K$ can be designed to be a robust controller. For example, if $\bm K$ is designed to be an ${\cal H}_\infty$ control for a given $\gamma>0$, then $\|z_1\|_{\cal P}< \gamma \|w\|_{\cal P}$. In general, for a given $\bm K$, $\|z_1\|_{\cal P}\le \|\bm {T_{z_1w}}(s)\|_\infty \|w\|_{\cal P}$. In this sense, the operator $\bm Q$ obtained in Proposition 1 achieves the two objectives of nominal tracking performance and minimization of $\|\bm {T_{z_1w}}(s)\|_\infty$ simultaneously. \pfbox
   \end{remark}

   \begin{remark}
       It is also noted that the controller $\bm C$ could itself be designed to address multiple optimal performance objectives such as the controller in Lemma 1 of \cite{vargas2013stabilization} that is derived from the Youla parameter to optimize multiple costs simultaneously. Since no impact of uncertain disturbance is counted, this controller could be complemented by the $\bm Q$ constructed in the MOCC design of this paper to achieve non-compromised robust optimal performance with respect to uncertain disturbance.  \pfbox
   \end{remark}

\begin{figure}[!ht]
			\centering
            \includegraphics[scale=0.7]{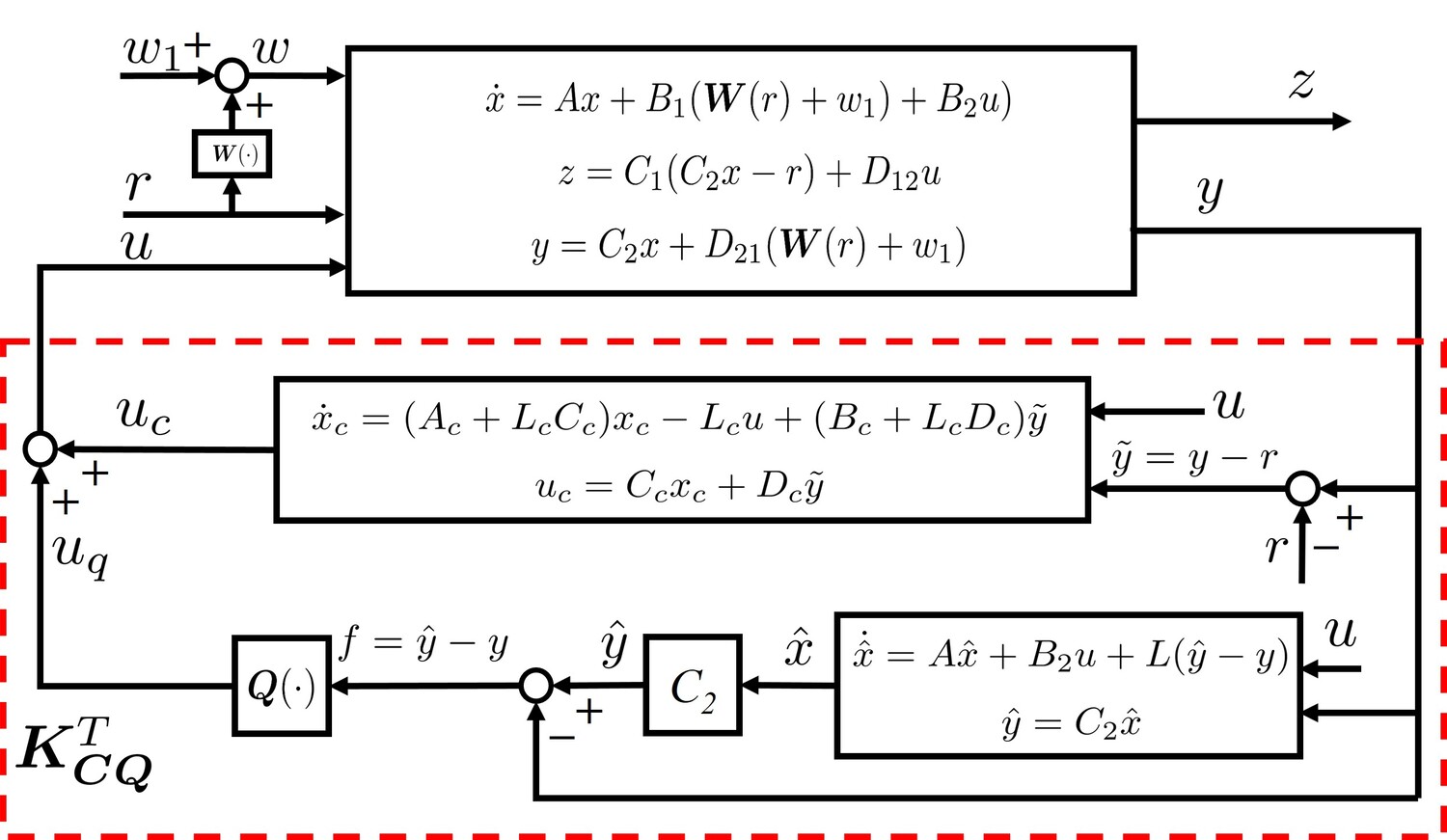}
			\caption{MO tracking control system by two-controller structure in the dependent case.}
			\label{fig:KCQ_tracking_dependence}
		\end{figure}

    The performance analysis for the dependent case could be more complicated due to the complexity of dependency between $w$ and $r$. In this paper, performance analysis is done for the case that $w=\bm W(r)+w_1$, where $\bm W (\cdot)$ is assumed to be a mapping satisfying $\|\bm W(r)\|_{\cal P}<\infty,\; \bm W(0)=0,\;\forall r\in {\cal P}$, and $w_1\in\mathcal{P}$ is the orthogonal part of $w$ with both $r$ and $\bm W(r)$, i.e., $\bm{S_{w_1r}}=0$ and $\bm{S_{w_1\bm W(r)}}=0$.  The system in (\ref{LTIsys:w}) then becomes
            \begin{align}
            \left\{ \mspace{-6mu}\begin{array}{ll}
					\dot x\mspace{-12mu}&=Ax+B_1(\bm W(r)+w_1)+B_2u\\
               z\mspace{-12mu}&=C_1(C_2x-r)+D_{12}u\\
    			y\mspace{-12mu}&=C_2x+D_{21}(\bm W(r)+w_1).
			\end{array} \right.
		\end{align}
     The MO tracking control system in Fig. \ref{fig:KCQ_tracking} can be modified as Fig. \ref{fig:KCQ_tracking_dependence}. 

    For a linear time-invariant mapping $\bm W(r)$, denoting the Laplace transform of $\bm W(r)$ as $\bm {W_r}(s)$, then
     \begin{align*}
       \|\bm W(r)\|_{\cal P}&=\sqrt{\frac{1}{2\pi}\int_{-\infty}^{\infty}{\rm Trace}\left[\bm{S_{\bm W(r)\bm W(r)}}(j\omega)\right]{\rm d}\omega}\\
       &=\|\bm {W_r}(s)\|_{\cal P},
     \end{align*}
     and $\bm w(s)=\bm {W_r}(s)+\bm w_1(s)$.
     Now we are ready to present the following performance analysis result in the dependent case.  
     \begin{theorem}\label{thm:dependent}
		For dependent $w$ and $r$ as in $w=\bm W(r)+w_1$, where $\bm W(r)$ is an LTI mapping with $\bm W_r(s)=\bm W(s)\bm r(s),  \bm W(s)\in\mathcal{RL}_{\infty}$,  and $w_1$ is orthogonal of both $r$ and $\bm W(r)$ in the sense that $\bm{S_{w_1r}}=0$ and $\bm{S_{w_1\bm W(r)}}=0$, given any controllers $\bm C$ and $\bm K$ in (\ref{C:tracking}) and (\ref{K:tracking}), the following equations hold:
			\begin{align}
            \|z\|_{\mathcal{P}}^2=& \| z_1\|_{\mathcal{P}}^2+\|\tilde z_2\|_{\mathcal{P}}^2,\label{robustp:dependent}\\
           \|z\|_{\mathcal{P}}^2-\gamma^2 \|w\|_\mathcal{P}^2=& \|z_1\|_{\mathcal{P}}^2-\gamma^2 \|w_1\|_\mathcal{P}^2 \notag\\
           &~~~~+ \|\tilde z_2\|_{\mathcal{P}}^2-\gamma^2 \|\bm W(r)\|_\mathcal{P}^2,\label{robustp:dependent_gamma}
			\end{align}
			where $\gamma$ is a scalar level constant,
        \begin{align*}
        \|z_1\|_{\mathcal{P}}^2&
        =\frac{1}{2\pi}\int_{-\infty}^{\infty}{\rm Trace}[\bm{T_{z_1w}}(j\omega)\bm{S_{w_1w_1}}(j\omega)\bm{T_{z_1w}}^*(j\omega)]{\rm d}\omega \\
        \bm{\tilde z_2}(s)&=\bm {T_{z_1w}}(s)\bm{W_r}(s)+\bm {T_{z_2r}}(s)\bm r(s)\\
        &=\bm {T_{z_1w}}(s)\bm{W_r}(s)+\bm{z_2}(s)\\
        \|z_2\|_\mathcal{P}^2&=\frac{1}{2\pi}\int_{-\infty}^{\infty}{\rm Trace}[\bm{T_{z_2r}}(j\omega)\bm{S_{rr}}(j\omega)\bm{T_{z_2r}^*}(j\omega)]{\rm d}\omega,  
            \end{align*}
            and $\bm {T_{z_1w}}(s)$ and $\bm {T_{z_2r}}(s)$ are derived in Lemma \ref{lem:cl}. Moreover, if $\bm K$ is designed to be an ${\cal H}_\infty$ control for a given $\gamma>0$ such that $\|\bm{T_{z_1w}}\|_{\infty}<\gamma$, then, regardless of $\bm C$, we have
        \begin{align}\label{worst_weighting}
            &\sup_{\bm{W(s)}\in\mathcal{RL}_{\infty}}\left\{\|\tilde z_2\|_{\mathcal{P}}^2-\gamma^2 \|\bm{W}(s)\bm r(s)\|_\mathcal{P}^2\right\}\notag\\
            =&\|\tilde z_2\|_{\mathcal{P}}^2-\gamma^2 \|\bm{\widetilde W}\bm r(s)\|_\mathcal{P}^2\notag\\
            =&\frac{1}{2\pi}\int_{-\infty}^{\infty}{\rm Trace}\big[\bm{S_{rr}}\bm{T_{z_2r}}^*(I-\gamma^{-2}\bm{T_{z_1w}}\bm{T_{z_1w}}^*)^{-1}\bm{T_{z_2r}}\big]{\rm d}\omega,
        \end{align}
        with the worst dependent weighting $\bm{\widetilde W}=(\gamma^2I-\bm{T_{z_1w}}^*\bm{T_{z_1w}})^{-1}\bm{T_{z_1w}}^*\bm{T_{z_2r}}$,  and
        \begin{align}\label{worst_performance}
            &\sup_{w\in\mathcal{P}}\left\{\|z\|_{\mathcal{P}}^2-\gamma^2 \|w\|_\mathcal{P}^2 \right\}=\sup_{w_1\in\mathcal{P}}\left\{ \|z_1\|_{\mathcal{P}}^2-\gamma^2 \|w_1\|_\mathcal{P}^2\right\} \notag\\
            &~~~~~~~~~~~~+ \sup_{\bm{W(s)}\in\mathcal{RL}_{\infty}}\left\{\|\tilde z_2\|_{\mathcal{P}}^2-\gamma^2 \|\bm{W}(s)\bm r(s)\|_\mathcal{P}^2\right\}\notag\\
            & \le \frac{1}{2\pi}\int_{-\infty}^{\infty} {\rm Trace}\big[ \bm{S_{rr}}\bm{T_{z_2r}}^*(I-\gamma^{-2}\bm{T_{z_1w}}\bm{T_{z_1w}}^*)^{-1}\bm{T_{z_2r}}\big]{\rm d}\omega.
        \end{align}  
		\end{theorem}
  
		\begin{IEEEproof}
            From Fig. \ref{fig:KCQ_tracking_dependence}, the closed-loop transfer function can be derived as:
            \begin{align*}
                \bm{z}(s)&=\bm {T_{z_1w}}(s)\bm w(s)+\bm {T_{z_2r}}(s)\bm r(s)\\
                        &=\bm {T_{z_1w}}(s)\bm{w_1}(s)+\bm {T_{z_1w}}(s)\bm{W_r}(s)+\bm {T_{z_2r}}(s)\bm r(s)\\
                &=\bm {z_1}(s)+\bm{\tilde z_2}(s), \\
                \bm{\tilde z_2}(s)&=\bm {T_{z_1w}}(s)\bm{W_r}(s)+\bm{z_2}(s), 
            \end{align*}
            noting that, if $r=0 \rightarrow \bm W(r)=0$, then $w=w_1\rightarrow \bm {T_{z_1w_1}}(s)=\bm {T_{z_1w}}(s)$, and $\bm {T_{z_1w}}(s)$ and $\bm {T_{z_2r}}(s)$ are derived in Lemma \ref{lem:cl}.  
     Since $\bm{S_{w_1r}}=0$ and $\bm{S_{w_1\bm W(r)}}=0$, we have
            \begin{align*}
                &\|z\|_\mathcal{P}^2=\frac{1}{2\pi}\int_{-\infty}^{\infty}{\rm Trace}[\bm{S_{zz}}(j\omega)]{\rm d}\omega\\
				=&\frac{1}{2\pi}\int_{-\infty}^{\infty}{\rm Trace}[\bm{T_{z_1w}}\bm{S_{w_1w_1}}\bm{T_{z_1w}^*}+\bm{S_{\tilde z_2\tilde z_2}}]{\rm d}\omega\\
                =&\|z_1\|_\mathcal{P}^2+\|\tilde z_2\|_\mathcal{P}^2.
            \end{align*}
    For \eqref{robustp:dependent_gamma}, it obviously holds, as
    \begin{align*}
        \|z\|_{\mathcal{P}}^2-\gamma^2 \|w\|_\mathcal{P}^2=& \|z_1\|_{\mathcal{P}}^2+ \|\tilde z_2\|_{\mathcal{P}}^2-\gamma^2 \|\bm {W}(r)+w_1\|_\mathcal{P}^2
    \end{align*}
    and $\bm{S_{w_1\bm W(r)}}=0$.
    
    For the LTI mapping with $\bm {W_r}(s)=\bm W(s)\bm r(s),\; \bm W(s)\in\mathcal{RL}_{\infty}$, since
    \begin{align*}
 \|\tilde z_2\|_{\mathcal{P}}^2-&\gamma^2 \|\bm W(s)\bm r(s)\|_\mathcal{P}^2\\
 &=\frac{1}{2\pi}\int_{-\infty}^{\infty}{\rm Trace}[\bm{S_{\tilde z_2\tilde z_2}}-\gamma^2\bm{W}\bm{S_{rr}}\bm{W}^*]{\rm d}\omega
\end{align*}
and
 \begin{align*}
            \bm{S_{\tilde z_2\tilde z_2}}=\left(\bm{T_{z_1w}}\bm{W}+\bm{T_{z_2r}}\right)\bm{S_{rr}}\left(\bm{T_{z_1w}}\bm{W}+\bm{T_{z_2r}}\right)^*,
        \end{align*}
    we have
 \begin{align*}
                &{\rm Trace}[\bm{S_{\tilde z_2\tilde z_2}}-\gamma^2\bm{W}\bm{S_{rr}}\bm{W}^*]\\
                =&{\rm Trace}\big[\bm{S_{rr}}\big(\bm{T_{z_2r}}^*\bm{T_{z_2r}}+\bm{W}^*\bm{T_{z_1w}}^*\bm{T_{z_2r}}+\bm{T_{z_2r}}^*\bm{T_{z_1w_1}}\bm W\\
                &+\bm{W}^*(\bm{T_{z_1w}}^*\bm{T_{z_1w}}-\gamma^2I)\bm{W}\big)\big]\\
                =&{\rm Trace}\big[\bm{S_{rr}}\big(\bm{T_{z_2r}}^*(I-\gamma^{-2}\bm{T_{z_1w}}\bm{T_{z_1w}}^*)^{-1}\bm{T_{z_2r}}\\
                &+(\bm{W}-\bm{\widetilde W})^*(\bm{T_{z_1w}}^*\bm{T_{z_1w}}-\gamma^2I)(\bm{W}-\bm{\widetilde W})\big)\big]
            \end{align*}
    with $\bm{\widetilde W}=(\gamma^2I-\bm{T_{z_1w}}^*\bm{T_{z_1w}})^{-1}\bm{T_{z_1w}}^*\bm{T_{z_2r}}$.
   Hence, it can be seen that, given arbitrary $w_1, r\in {\cal P}$, $\sup_{\bm{W(s)}\in\mathcal{RL}_{\infty}}\left\{\|z\|_{\mathcal{P}}^2-\gamma^2 \|w\|_\mathcal{P}^2\right\}$ is achieved at $\bm W=\bm{\widetilde W}$ since $(\bm{T_{z_1w}}^*\bm{T_{z_1w}}-\gamma^2I)<0$, which renders the worst dependent weighting $\bm{\widetilde W}$. Therefore, \eqref{worst_weighting} is proved and  \eqref{worst_performance} is straightforward by noting the inequality 
   \begin{align*}
    \|z_1\|_{\mathcal{P}}^2\le \|\bm{T_{z_1w}}\|_\infty^2 \|w_1\|_{\mathcal{P}}^2< \gamma^2\|w_1\|_{\mathcal{P}}^2.
   \end{align*}
		\end{IEEEproof}

        \begin{remark}
            If no dependence is assumed between $w$ and $r$, that is $\bm W(r)=0$, 
          then $\tilde z_2=z_2$, thus, resulting in Theorem \ref{thm:orthogonal}. \pfbox
        \end{remark}

        \begin{remark}\label{rem:performance_dependent}
            From Theorem \ref{thm:dependent}, one can derive
            \begin{align*}
            &\|z\|^2_{\cal P}=\|z_1\|^2_{\cal P}+\|\tilde z_2\|^2_{\cal P}\\
            =&\|\bm {T_{z_1w}}(s)\bm{w_1}(s)\|^2_{\cal P}+\|\bm {T_{z_1w}}(s)\bm {W_r}(s)+\bm {T_{z_2r}}(s)\bm r(s)\|^2_{\cal P}\\
            \le &\|\bm {T_{z_1w}}(s)\|_{\infty}^2 (\|w_1\|_{\cal P}^2+\|\bm {W}(r)\|_\mathcal{P}^2)+\|z_2\|_{\mathcal{P}}^2\\
            = &\|\bm {T_{z_1w}}(s)\|_{\infty}^2 \|w\|_{\cal P}^2+\|z_2\|_{\mathcal{P}}^2,
            \end{align*} 
            where the fact that $\|\bm {W}(r)\|_\mathcal{P}=\|\bm {W_r}(s)\|_\mathcal{P}$ is applied. 
            It can be seen that the upper bound above is a sum of two terms that depend on controllers $\bm K$ and $\bm C$ separately. Hence, it is clear that, for any $\|w_1\|_{\cal P}<\infty$ and $\|\bm W(r)\|_{\cal P}<\infty$, the same observation as that from Theorem \ref{thm:orthogonal} can be drawn: to achieve better robust tracking performance, ${\bm C}$ should be designed to bring the nominal tracking performance $\|z_2\|_{\cal P}$ low and ${\bm K}$ should be designed to address the robustness against $w=\bm W(r)+w_1$. It should be pointed out that the dependency $\bm W(r)$ may not be always harmful to the tracking performance, since the case $\|\bm {T_{z_1w}}(s)\bm {W_r}(s)+\bm {T_{z_2r}}(s)\bm r(s)\|^2_{\cal P}<\|\bm {T_{z_2r}}(s)\bm r(s)\|^2_{\cal P}$ could happen. In case it poses a negative impact, Theorem \ref{thm:dependent} gives the worst possible linear dependency $\bm W(s)=\bm{\widetilde W}(s)$ for an ${\cal H}_\infty$ controller $\bm K$.\pfbox
            \end{remark}

		\subsection{A specific MOCC design: LQ tracking (LQT) $+\mathcal{H}_{\infty}$}\label{subsec:LQT+Hinf}
		In this section, a specific MO complementary tracking controller is presented based on the performance analysis results in Theorems \ref{thm:orthogonal} and \ref{thm:dependent}. That is, controllers ${\bm C}$ and ${\bm K}$ are designed according to the LQ optimal criterion and the $\mathcal{H}_{\infty}$ criterion, respectively. Besides Assumption \ref{ass:AB2C2}, the following standard assumption is made on the system (\ref{LTIsys:w}) in this subsection \cite{glover1988state,zhou1996robust}:\\
        \begin{assumption}
            (i) $\left[\begin{array}{cc}
		  A-j\omega I & B_1\\
    		C_2 & D_{21} 
	   \end{array}\right]$ has full row rank for all $\omega$; (ii) $\left[\begin{array}{cc}
		A-j\omega I & B_2\\ 
		C_1C_2 & D_{12}
	   \end{array}\right]$ has full column rank for all $\omega$; (iii) $R_1:=D_{12}'D_{12}>0$ and $R_2=D_{21}D_{21}'>0$.
        \end{assumption}

		\emph{Step 1:  Design controller ${\bm C}$ to be the LQ optimal tracking controller.} 
		The design of the LQ optimal tracking controller is based on the FDLTI system (\ref{LTIsys:w}) with $w=0$, i.e., system (\ref{LTIsys:w0}) and, is to minimize the following quadratic cost function:
		\begin{align}\label{sys:cost}
			J=\|z_2\|_{\mathcal P}^2=\lim\limits_{T\to\infty}\frac{1}{T}\int_{0}^{T} \|z_2\|^2 {\rm d}t.
		\end{align}
        The optimal control law, which is a 2DoF form\footnote{Though the tracking controller $\bm C$ in (\ref{C:tracking}) is in the 1DoF form, there is no restriction for the proposed MOCC framework to be applied to the case of a 2DoF controller $\bm C$, as discussed in Remark \ref{rem:2dof}.}, is given by \cite[Chapter 4]{anderson1989optimal}\cite{shaked1995continuous}
        \begin{align}\label{LQT}
					\left\{ \mspace{-6mu}\begin{array}{l}
				\dot{\hat x} =A\hat x+B_2u+L(C_2\hat x-y)\\
				\dot b=-(A+B_2F)'b+Sr\\
				u= F\hat x-R_1^{-1}B_2'b-R_1^{-1}D_{12}'C_1r
			\end{array} \right.
		\end{align}
		and the minimal cost is
				\begin{align}\label{Joptimal}
					J_*=\lim\limits_{T\to\infty}\frac{1}{T}\int_{0}^{T} c(\tau) {\rm d}\tau,
				\end{align}
		where $L$ is the observer gain matrix with $A+LC_2$ stable, $F=-R_1^{-1}(\Pi B_2+C_2'C_1'D_{12})'$, $S=C_2'C_1'C_1-(\Pi B_2+C_2'C_1'D_{12})R_1^{-1}D_{12}'C_1$, and $c(\tau)=\|C_1r(\tau)\|^2-\|R_1^{-1/2}(B_2'b(\tau)-D_{12}'C_1r(\tau))\|^2$. Here, $\Pi\ge 0$ is the stabilizing solution to the following algebraic Riccati equation:
			\begin{align*}
				\Pi A+A'\Pi
				-(\Pi B_2+C_2'C_1'D_{12})R_1^{-1}&(\Pi B_2+C_2'C_1'D_{12})' \notag\\&+ C_2'C_1'C_1C_2=0.
			\end{align*}
		Notice that the dynamics of $b(t)$ in (\ref{LQT}) is anticausal such that $b(t)$ is bounded \cite[Chapter 4]{anderson1989optimal}.
		\emph{Step 2: Design controller ${\bm K}$ to be an $\mathcal{H}_{\infty}$ controller.} Now consider the FDLTI system (\ref{LTIsys:w}) with $r=0$, i.e., system (\ref{LTIsys:r0}). The $\mathcal{H}_{\infty}$ controller is designed according to the following performance criterion:
		\begin{align}\label{hinf}
			\|\bm{T_{z_1w}}(s)\|_{\infty}<\gamma,
		\end{align}
		where $\gamma>0$ is a prescribed value.
        It follows from \cite{zhou1996robust,glover1988state} that the central $\mathcal{H}_{\infty}$ controller satisfying (\ref{hinf}) is 
		\begin{align}\label{controller:hinf}
			\dot x_{\infty}&=A_{\infty}x_{\infty}+B_{\infty}y,\notag\\
			 u_{\infty}&=C_{\infty}x_{\infty},
		\end{align}
		where 
		\begin{align*}
			A_{\infty}&=A+\gamma^{-2}B_1B_1'P_1+B_2C_\infty-B_{\infty}(C_2+\gamma^{-2}D_{21}B_1'P_1)\notag\\
	    B_{\infty}&=-(I-\gamma^{-2}P_1P_2)^{-1}L_{\infty}\notag\\
	C_{\infty}&=-R_1^{-1}(P_1B_2+C_1'D_{12})'\notag\\
	L_{\infty}&=-(C_2P_2+D_{21}B_1')'R_2^{-1}.
		\end{align*}
		Here, $P_1\ge 0$ and $P_2\ge 0$ are the stabilizing solutions to the following algebraic Riccati equations:
		\begin{align*}
			P_1A+&A'P_1+\gamma^{-2}P_1B_1B_1'P_1+C_2'C_1'C_1C_2\notag\\
			&-(P_1B_2+C_2'C_1'D_{12})R_1^{-1}(P_1B_2+C_2'C_1'D_{12})'=0,\notag\\
			P_2A'+&AP_2+\gamma^{-2}P_2C_1'C_1P_2+B_1B_1'\notag\\
			&~~~~~~~-(C_2P_2+D_{21}B_1')'R_2^{-1}(C_2P_2+D_{21}B_1')=0.
		\end{align*}
        
		\emph{Step 3:  Design of MOCC: ${\rm LQT}+\mathcal{H}_{\infty}$  controller.}
		
\begin{figure}[!ht]
			\centering
			\includegraphics[scale=0.35]{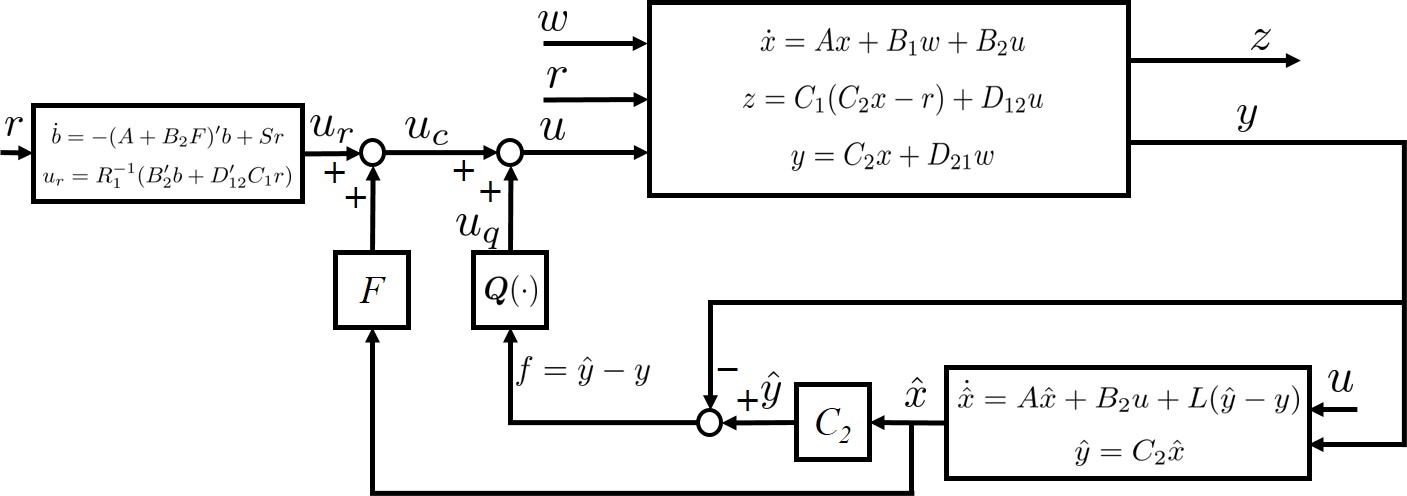}
			\caption{Diagram of ${\rm LQT}+\mathcal{H}_{\infty}$  controller by the MOCC framework.}
			\label{fig:LQT_Hinf}
		\end{figure}

		Applying the two-controller structure with shared observer (Proposition \ref{prop:sharedQ}) gives the following MO complementary tracking controller ${\rm LQT}+\mathcal{H}_{\infty}$:
		\begin{eqnarray}\label{KCQLQT:dynamic}
			\left\{ \mspace{-6mu}\begin{array}{l}
				\dot{\hat x} =A\hat x+B_2u+Lf\\
				\dot b=-(A+B_2F)'b+Sr\\
				u_c= F\hat x-R_1^{-1}B_2'b-R_1^{-1}D_{12}'C_1r\\
				\dot x_q=A_qx_q+B_qf\\
				u_q=C_qx_q+D_qf\\
				u=u_c+u_q,
			\end{array} \right.
		\end{eqnarray}
		where $f=C_2\hat x-y$, and	by (\ref{Qshared:ABCD}) in Proposition \ref{prop:sharedQ},
		\begin{align*}
			A_q&=\left[\begin{array}{cc}
					A & B_2C_{\infty} \\ 
					B_{\infty}C_2 & A_{\infty}
				\end{array}\right],\;
				 B_{q}=\left[\begin{array}{c}
					L\\ 
					-B_{\infty}
				\end{array}\right],\\
				C_q&=\left[\begin{array}{cc}
					-F & C_{\infty}
				\end{array}\right],\;
                D_q=0.
		\end{align*}
		The diagram of ${\rm LQT}+\mathcal{H}_{\infty}$  controller by the MOCC framework is shown in Fig. \ref{fig:LQT_Hinf}. Similarly, by applying the two-controller structure in the state-feedback case in subsection \ref{subsec:static}, the ${\rm LQT}+\mathcal{H}_{\infty}$  controller of the state-feedback version can be obtained.
		
		Finally, for the MO complementary tracking controller (\ref{KCQLQT:dynamic}), it follows from Remarks \ref{rem:performance_orthogonal} and \ref{rem:performance_dependent} that the performance measure $\|z\|_\mathcal{P}^2$ can be evaluated by
		\begin{align}\label{robustp:LQT_Hinfty}
			\|z\|_\mathcal{P}^2\le J_*+\|\bm{T_{z_1w}}\|_\infty^2\|w\|_\mathcal{P}^2,
		\end{align}
		where $J_*$ is the minimal cost in the LQ optimal tracking given by (\ref{Joptimal}). 
		
		\begin{remark}
			Note that the controller ${\bm K}$ can also be designed as a mixed  $\mathcal{H}_{2}/\mathcal{H}_{\infty}$ controller \cite{chen2001multiobjective,bernstein1989lqg} if stochastic noises need to be addressed.
		\end{remark}
		
		
		\section{Data-driven Optimization of Performance}\label{sec:optimization}
		Theorems  \ref{thm:orthogonal} and \ref{thm:dependent} show the total tracking performance achieved by $\bm C$ and 
   ${\bm K}$.  Note that, if ${\bm K}$ is designed as the $\mathcal{H}_{\infty}$ controller shown in Subsection \ref{subsec:LQT+Hinf},  it is still the worst-case design. Hence, in order to further improve the robust performance, an extra gain factor $\alpha\in \mathbb{R}$ can be introduced into the operator ${\bm Q}$:
		\begin{align}\label{Qalpha}
			\bm{Q_{\alpha}}:\left\{ \mspace{-6mu}\begin{array}{l}
				\dot x_q=A_qx_q+B_qf\\
				u_q=\alpha C_qx_q+\alpha D_qf,
			\end{array} \right.
		\end{align}
		with $(A_q, B_q, C_q, D_q)$ given in (\ref{Qgeneral:ABCD}). This $\alpha$ can be further optimized through a data-driven approach. Let us denote the MO tracking controller with 
  $\bm{Q_{\alpha}}$ 
  as $\bm{K_{CQ_{\alpha}}^{T}}$. Apparently, the two special cases $\alpha=0$ and $\alpha=1$ correspond to the nominal controller $\bm C$ and the MO complementary controller $\bm{K_{CQ}^T}$, respectively.
		
  With $\left[\begin{array}{cccc}
			x'  & x_c' & x_q' & x'-\hat x'
		\end{array}\right]'$ being the state of the closed-loop system, the closed-loop system matrix becomes
		\begin{align*}
			\bar A(\alpha)=\left[\mspace{-6mu}\begin{array}{cccc}
				A+B_2D_cC_2 & B_2C_c & \alpha B_2C_q & -\alpha B_2D_qC_2  \\
				B_cC_2 & A_c & -\alpha L_cC_q & \alpha L_cD_qC_2  \\
				0   & 0 & A_q & B_qC_2 \\
                0  & 0 & 0 & A+LC_2\\
			\end{array}\mspace{-6mu}\right].
		\end{align*}
		Obviously, $\bar A(\alpha)$ is a stable matrix for all $\alpha\in \mathbb{R}$,
        so that the stability of the closed-loop is guaranteed for any $\alpha\in \mathbb{R}$. The system performance can be further improved by $\alpha$ through a data-driven approach. In this regard, let us consider the task of tracking a desired trajectory within a specified finite time interval, with the task repeating over multiple iterations. It is assumed that the disturbance is iteration-invariant; in other words, when the tracking task is repetitive, the unknown disturbance also repeats. This scenario closely resembles the setting in \cite{killingsworth2006pid}, where PID parameters are iteratively tuned.
        Thus, the performance optimization problem can be dealt with in the framework of iterative learning control (ILC) \cite{bristow2006survey,khong2016iterative}, and the following finite-horizon cost function is introduced:
		\begin{align}\label{sys:cost_alpha}
			J(\alpha)=\frac{1}{T}\int_{0}^{T} \|z_m(t)\|^2 {\rm d}t,
		\end{align}
		where $z_m=C_1(y-r)+D_{12}u$ is the measured version of $z$.
		Then an optimal $\alpha$  minimizing the cost function (\ref{sys:cost_alpha}) can be found by an extremum seeking (ES) algorithm \cite{ariyur2003real,tan2006non,killingsworth2006pid}, which requires the following assumption. 
		
		\begin{assumption}\label{ass:es}
		    For the closed-loop system with the controller $\bm{K_{CQ_{\alpha}}^{T}}$ and in the presence of a deterministic and repeating disturbance $w(t), t\in [0,T]$,  the cost function defined by (\ref{sys:cost_alpha}) has a minimum at $\alpha=\alpha^*$, and the following holds:
			\begin{align}
				\frac{\partial J(\alpha)}{\partial \alpha}\bigg |_{\alpha=\alpha^*}=0, \; \frac{\partial^2 J(\alpha)}{\partial \alpha^2}\bigg |_{\alpha=\alpha^*}>0.
			\end{align}
		\end{assumption}
		
\begin{figure}[!ht]
			\centering
            \includegraphics[scale=0.4]{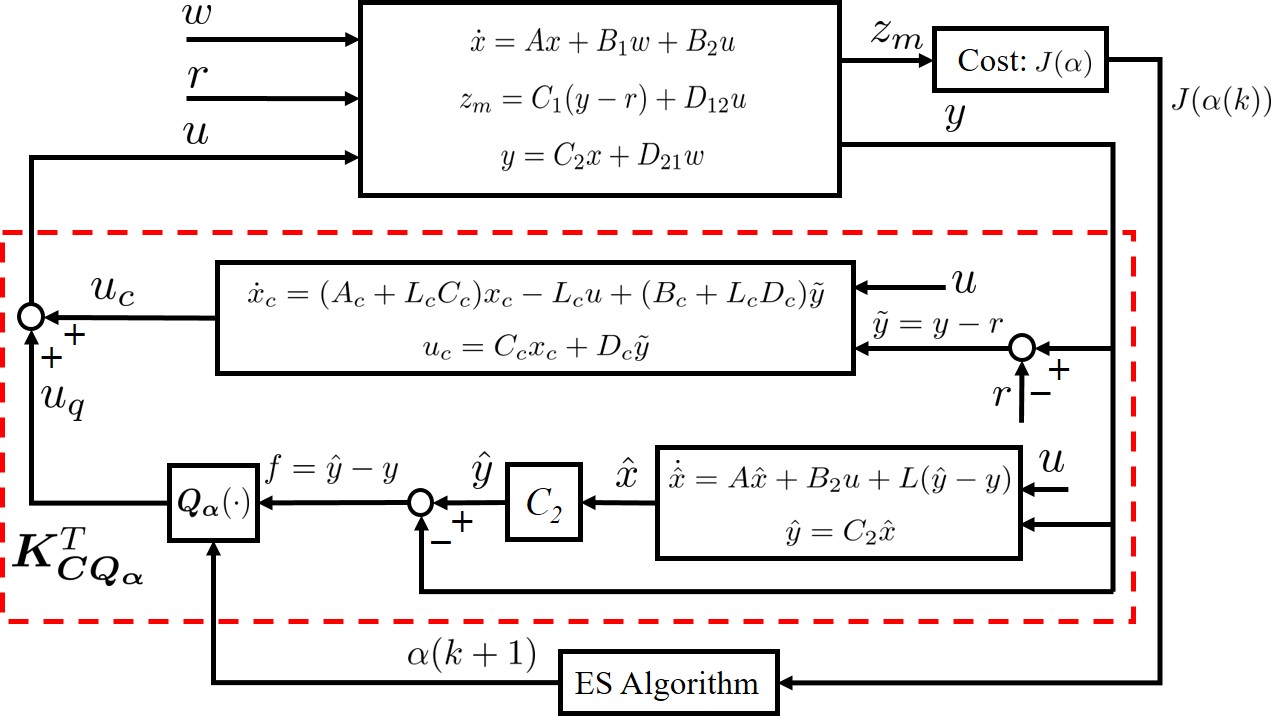}
			\caption{The ES $\alpha$ tuning diagram: the ES algorithm updates the parameter $\alpha(k)$ in $\bm{Q_{\alpha}}$ iteratively to minimize $J(\alpha)$.}
			\label{fig:ES}
		\end{figure}

		Under Assumption \ref{ass:es},  an ES algorithm can be used to tune the parameter $\alpha$ for a given disturbance $w(t)$ by repeatedly running the closed-loop system with the controller $\bm{K_{CQ_{\alpha}}^{T}}$ over the finite time $[0,T]$.
		ES is a data-driven optimization method which uses input-output data to seek an optimal input with respect to a selected cost \cite{ariyur2003real}. The ES algorithm adopted here is in the iteration domain, which can be, for example, that presented in \cite{killingsworth2006pid}. 
		The overall ES tuning scheme for $\alpha$  is delineated in Fig. \ref{fig:ES}. By tuning the parameters of the ES algorithm appropriately, the parameter $\alpha$ will converge to a small neighborhood of $\alpha^*$ iteratively as shown in our work \cite{xu2023robust}.

		\section{Example}\label{sec:example}
		To demonstrate the advantages of the developed MO control framework, a comparative example of tracking control is worked out. Consider the following double integrator system
    \begin{align*}
			\dot x&=\left[\begin{array}{cc}
		0 & 1\\
		0 & 0
	\end{array}\right]x+\left[\begin{array}{c}
	1 \\ 10
	\end{array}\right]w+\left[\begin{array}{c}
		0 \\
		1
	\end{array}\right]u,\notag\\
			y&=\left[\begin{array}{cc}
		1 & 0
	\end{array}\right]x+0.01w,
		\end{align*}
		with the performance variable
		\begin{align*}
			z=\left[\begin{array}{c}
		1 \\ 0
	\end{array}\right](\left[\begin{array}{cc}
		1 & 0
	\end{array}\right]x-r)+\left[\begin{array}{c}
		0 \\ 0.03
	\end{array}\right]u.
		\end{align*}
    The reference signal is $r=\sin(\pi t)$. Thus, the tracking task is to design a controller such that the first state $x_1$ can track the reference signal $r$ while the control effort is taken into account.  To make comparisons, four different tracking control methods are considered: the proposed MOCC in (\ref{KCQLQT:dynamic}), LQT control in (\ref{LQT}), $\mathcal{H}_{\infty}$ tracking control \cite{shaked1995continuous}, and disturbance observer-based control (DOBC) \cite{chen2016disturbance}. For the MO complementary controller (\ref{KCQLQT:dynamic}), 
    \begin{align*}
        F=\left[\begin{array}{cc}
		-33.33 &-8.17
	\end{array}\right],\; L=\left[\begin{array}{cc}
		-100 &-1000
	\end{array}\right]',
    \end{align*}
    and the robustness level is chosen as $\gamma=0.4108$, which is the minimum $\gamma$ value (tolerance of $10^{-4}$). The LQT controller uses the same $F$ and $L$, and the $\mathcal{H}_{\infty}$ tracking controller uses the same $\gamma$.
    Here, the DOBC combines a disturbance observer and a disturbance compensation gain, whose structure is similar to the proposed two-controller structure with shared observer in Fig. \ref{fig:KCQ_shared}. The disturbance observer is to jointly estimate the system state and disturbance:
    \begin{align*}
        &\left\{ \mspace{-6mu}\begin{array}{l}
				\dot{\hat \chi} =A\hat \chi+B_1\hat w+B_2u+L_{\chi}(\hat y-y)\\
				\hat y=C_2\hat \chi+D_{21}\hat w,
			\end{array} \right.\\
        &\left\{ \mspace{-6mu}\begin{array}{l}
				\dot{\hat \xi} =A_w\hat \xi +L_w(\hat y-y)\\
				\hat w=C_w\hat\xi,
			\end{array} \right.
    \end{align*}
    where $\hat\chi$ and $\hat w$ are the estimated state and disturbance, respectively, and $A_w$ and $C_w$ are the preassumed disturbance model. The selection of $L_x, L_w, A_w$, and $C_w$ needs to make the error matrix $\left[\begin{array}{cc}
		A+L_xC_2 & B_1C_w+L_{\chi}D_{21}C_w\\
		L_wC_2 & A_w+L_wD_{21}C_w
	\end{array}\right]$ stable. The disturbance compensation gain can be chosen as \cite{chen2016disturbance,yang2011robust}
    \begin{align*}
        F_w=-[C_2(A+B_2F)^{-1}B_2]^{-1}C_2(A+B_2F)^{-1}B_1.
    \end{align*}
    Then the tracking controller based on the DOBC method is designed to be $u=F\hat\chi+F_w\hat w-R_1^{-1}B_2'b-R_1^{-1}D_{12}'C_1r$.

\begin{figure}[!ht]
			\centering
			\includegraphics[scale=0.55]{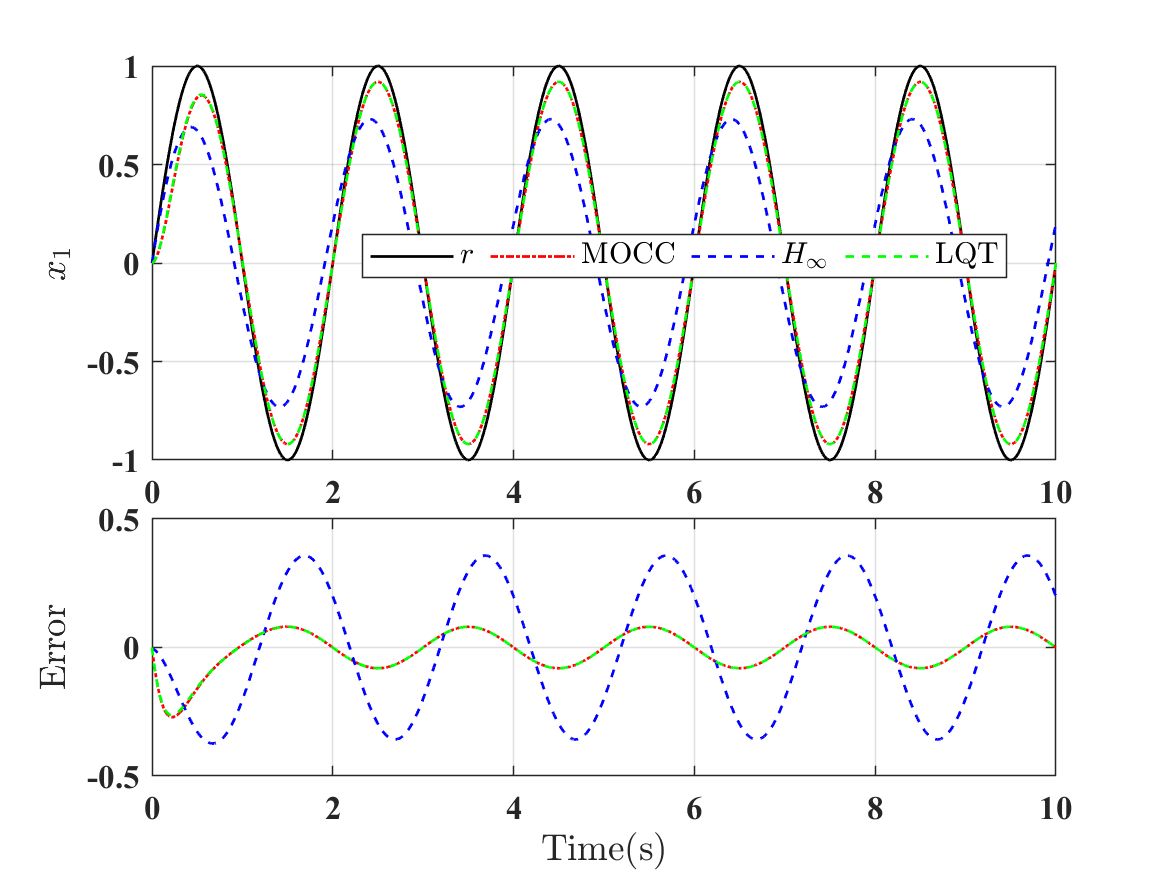}
			\caption{Tracking response and tracking error in the absence of disturbance.}
			\label{fig:tracking_response}
		\end{figure}

    First, we consider the scenario in the absence of disturbance. In this scenario, the DOBC is not  included for comparison as no disturbance is considered. The tracking response and tracking error are shown in Fig. \ref{fig:tracking_response} and the average  quadratic costs $\frac{1}{T}\int_{0}^{T} \|z(t)\|^2 {\rm d}t$ $(T=100s)$ for MOCC, LQT control, and $\mathcal{H}_{\infty}$ tracking control are  $0.0408, 0.0408$, and $0.1127$, respectively. Thus, it is verified that the MOCC has the same tracking performance as LQT control when $w=0$, while, on the other hand, it is shown that the $\mathcal{H}_{\infty}$ control has a significant performance loss ($176\%$) compared with the LQT control. 

\begin{figure}[!ht]
			\centering
			\includegraphics[scale=0.55]{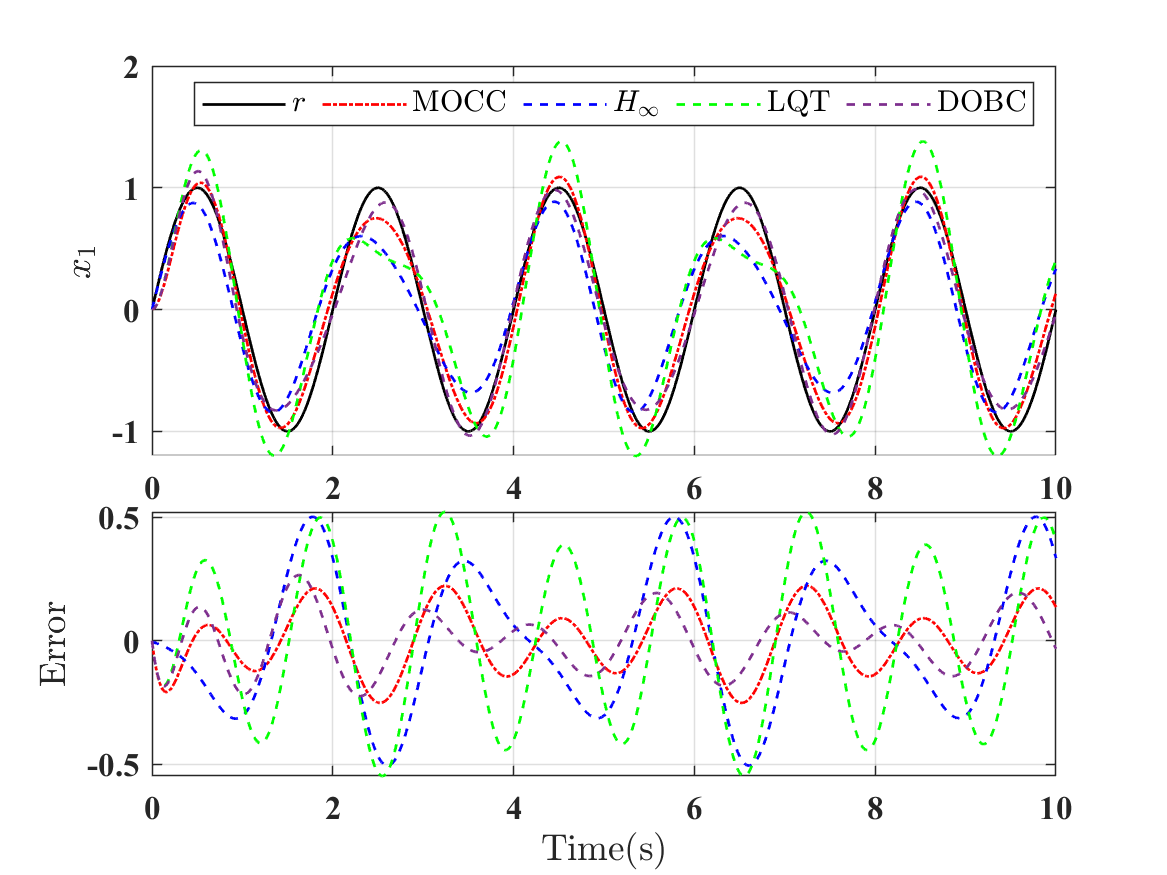}
			\caption{Tracking response and tracking error in the presence of $w_1$.}
			\label{fig:tracking_response_w1}
		\end{figure}

\begin{figure}[!ht]
			\centering
			\includegraphics[scale=0.55]{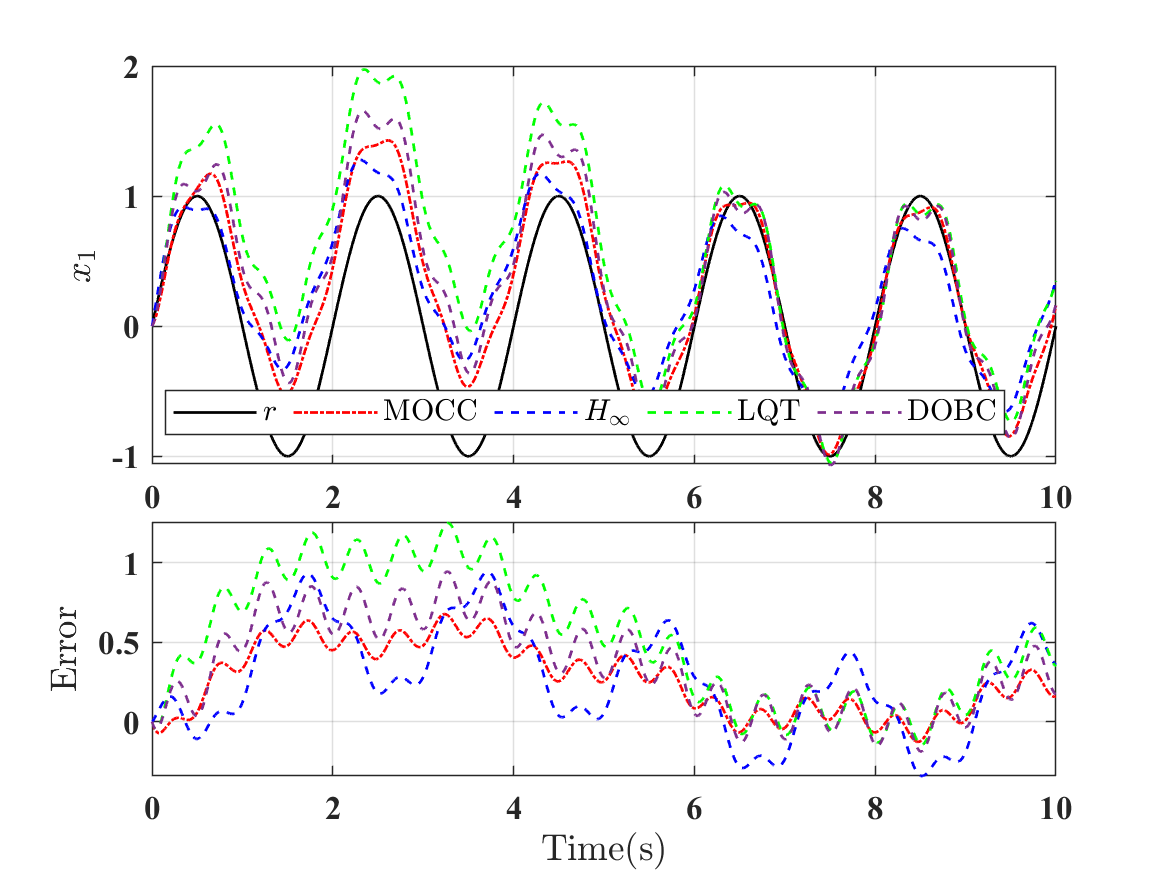}
			\caption{Tracking response and tracking error in the presence of $w_2$.}
			\label{fig:tracking_response_w2}
		\end{figure}

\begin{figure}[!ht]
			\centering
			\includegraphics[scale=0.55]{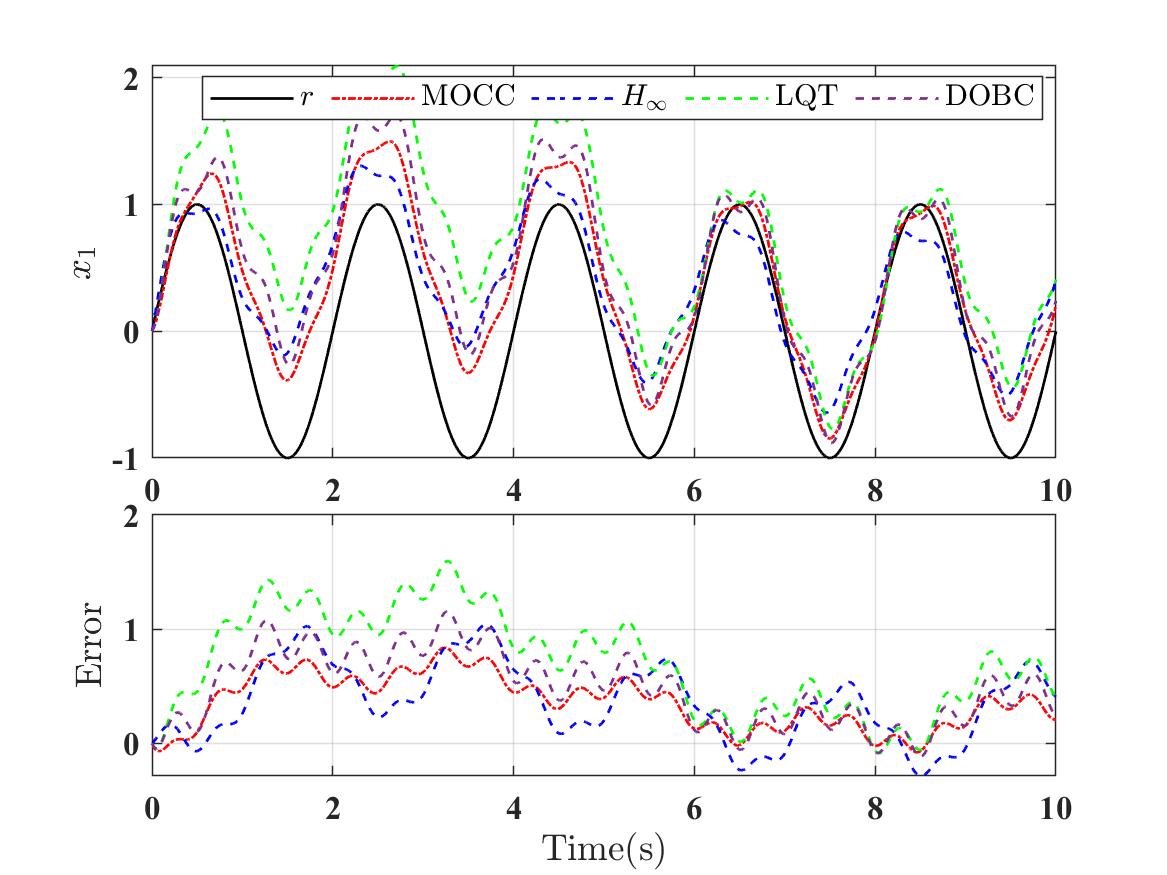}
			\caption{Tracking response and tracking error in the presence of $w_3$.}
			\label{fig:tracking_response_w3}
		\end{figure}

    Now three different disturbances are considered: 
    \begin{align*}
        w_1&=\sin(1.5\pi t),\\
        w_2&=\sin(4\pi t)+\sin(0.2\pi t)+1,\\
        w_3&=W(r)+w_2,\; \dot W(r)=0.01W(r)+r.
    \end{align*}
    For the DOBC method, the matrices $A_w$ and $C_w$ use the model of $w_1$, i.e., $A_w=\left[\begin{array}{cc}
		0 & 1\\
		-2.25\pi^2 & 0
	\end{array}\right], C_w=\left[\begin{array}{cc}
		1 & 0
	\end{array}\right]$. Choosing $L_x=L$ and $L_w=\left[\begin{array}{cc}
		-200 & -1000
	\end{array}\right]'$ such that the eigenvalues of the error matrix $\left[\begin{array}{cc}
		A+L_xC_2 & B_1C_w+L_{\chi}D_{21}C_w\\
		L_wC_2 & A_w+L_wD_{21}C_w
	\end{array}\right]$ are $(-88.73, -11.27,  -1 \pm 5.59i)$. The tracking results for $w_1, w_2, w_3$ are shown in Figs. \ref{fig:tracking_response_w1}--\ref{fig:tracking_response_w3}. It can be observed in Fig. \ref{fig:tracking_response_w1} that the DOBC is comparable to MOCC in terms of tracking error in the presence of $w_1$. This is not surprising, as the DOBC uses the exact disturbance model of $w_1$. We can calculate the quadratic costs $\frac{1}{T}\int_{0}^{T} \|z\|^2 {\rm d}t$ ($T=100s$) generated by the four controllers for $w_1$, $w_2$, and $w_3$. Also, the robust performance of the four controllers can be compared by computing the $\mathcal{H}_{\infty}$ norm of the closed-loop transfer matrix from $w$ to $z$. The quadratic cost and the $\mathcal{H}_{\infty}$ norm for the four controllers are summarized in Table I. It can be seen that the proposed MOCC method generates the minimum tracking cost for the three specific disturbances and has the minimum $\mathcal{H}_{\infty}$ norm for robustness. In summary, the proposed MOCC method does not require \emph{a prior} knowledge of the disturbance model and guarantees a certain robustness level $\gamma$ compared to the LQT and the DOBC, and performs better than the $\mathcal{H}_{\infty}$ method due to the proposed complementary structure. 
       \begin{table}[!htbp]\label{table1}
		\small
		\renewcommand{\arraystretch}{1.3}
		\centering 
		\caption{Performance comparison for different controllers in terms of cost $\frac{1}{T}\int_{0}^{T} \|z\|^2 {\rm d}t$ ($T=100s$).}
		\begin{tabular}{|c|cccc|}
			\hline
			& MOCC & $\mathcal{H}_{\infty}$ & LQT & DOBC \\ \hline
			$w=0$ & $\bm{0.0408}$ & $0.1127$
			& $0.0408$ & $0.0408$\\ \hline
            $w_1$ & $\bm{0.1245}$ & $0.1970$
			& $0.2629$ & $0.1319$\\ \hline
            $w_2$ & $\bm{0.3771}$ & $0.4499$
			& $0.6698$ & $0.4744$\\ \hline
            $w_3$ & $\bm{0.6296}$ & $0.6946$
			& $1.2425$ & $0.8159$\\ \hline
			$\mathcal{H}_{\infty}$ norm & $\bm{0.4108}$ & $0.4108$ & $0.6835$
			& $1.0020$ \\ \hline
		\end{tabular}
	\end{table}


\begin{figure}[!ht]
			\centering
			\includegraphics[scale=0.5]{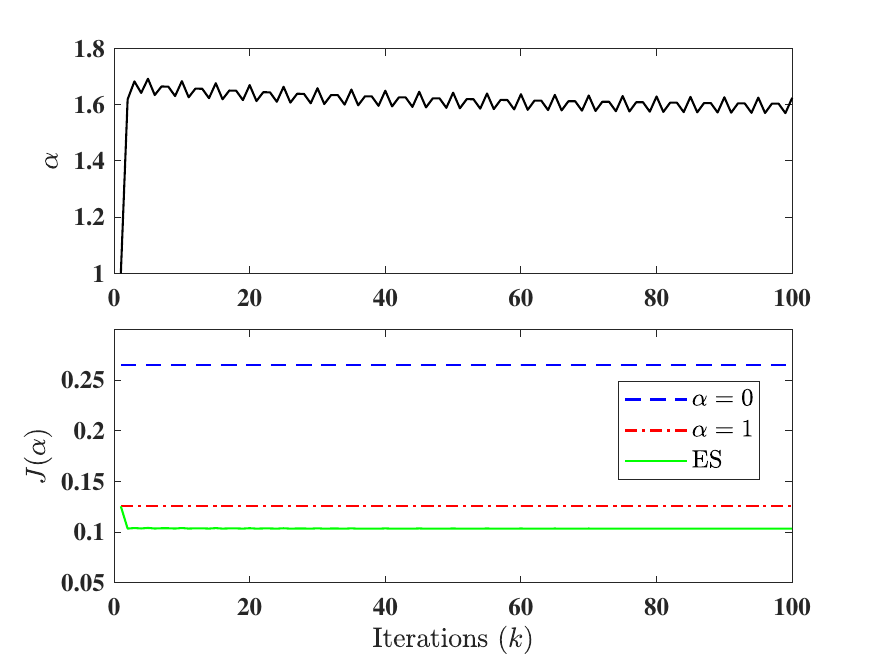}
			\caption{$\alpha$ tuning in the presence of $w_1.$}
			\label{fig:alpha_simulation}
		\end{figure}

    Finally, we provide a simulation to illustrate the performance optimization idea with a factor $\alpha$ discussed in Section \ref{sec:optimization}.   We will use the ES algorithm in the iteration domain from \cite{killingsworth2006pid} to find an optimal $\alpha$. The quadratic cost $J(\alpha)$ in \eqref{sys:cost_alpha} is used with $T=100s$. The disturbance signal $w_1=\sin(1.5\pi t)$ is considered in the $\alpha$ tuning process and the result is shown in Fig. \ref{fig:alpha_simulation}. It is seen that  the parameter $\alpha$ converges to a small neighborhood of $\alpha^*=1.60$ which yields a lower cost $J(\alpha=\alpha^*)=0.1035$ compared with the cases when  $\alpha=0$ ($J(\alpha=0)=0.2654$, LQT performance) and $\alpha=1$ ($J(\alpha=1)=0.1256$, MOCC performance).

		\section{Conclusion}\label{sec:conclusion}
	    A multi-objective complementary control (MOCC) framework that can assemble two independently designed controllers in a complementary way instead of a trade-off 
        is proposed. A state-space realization for the Youla-type operator ${\bm Q}$ is provided to manage the two controllers. In particular, an MO complementary tracking control is applied to demonstrate the advantages of MOCC. Rigorous performance analysis shows that the tracking performance and robustness can be addressed separately, especially when the disturbance signal is independent of the reference signal. Simulation results validate the advantages of MOCC, especially, when the disturbance signal is completely unknown. Furthermore, it is shown that this framework can be potentially extended to improve the total performance through a data-driven approach with an extra gain factor to $\bm Q$. 

	
		\bibliographystyle{IEEEtran}
		\bibliography{mybibfile}

	\end{document}